\documentclass{amsart}

\usepackage{amsmath}    
\usepackage{amssymb}    
\usepackage{dsfont} 	
\usepackage{bm}         
\usepackage{mathtools}  
\usepackage[hypertexnames=false]{hyperref} 
\usepackage[scientific-notation=true]{siunitx}
\usepackage{enumerate}
\usepackage{centernot}
\usepackage{pdflscape}
\usepackage{paralist}
\usepackage{stackrel}
\usepackage[english]{babel}
\usepackage[table]{xcolor}
\usepackage{setspace}

\usepackage[numbers]{natbib}
\usepackage{bibunits}

\newcommand*{\QEDE}{\hfill\ensuremath{\blacksquare}}
\def\ci{\perp\!\!\!\perp}

\usepackage{algorithm}
\usepackage{algpseudocode}

\usepackage{listings}
\usepackage{color}

\newtheorem*{heuristic*}{Heuristic Definition}
\theoremstyle{definition}
\newtheorem{Example}{Example}

\makeatletter
\newcommand{\addresseshere}{%
  \enddoc@text\let\enddoc@text\relax
}

\definecolor{lightgrey}{rgb}{0.9,0.9,0.9}
\definecolor{darkgreen}{rgb}{0,0.6,0}
\begin{document}


\title[The Spectral Condition Number Plot]
{The Spectral Condition Number Plot for Regularization Parameter Determination}

\author[C.F.W.\ Peeters]{Carel F.W.\ Peeters*}
\thanks{*Principal corresponding author.}
\address[Carel F.W.\ Peeters]{
Dept.\ of Epidemiology \& Biostatistics \\
VU University medical center Amsterdam \\
Amsterdam\\
The Netherlands}
\email{cf.peeters@vumc.nl}

\author[M.A.\ van de Wiel]{Mark A.\ van de Wiel}
\address[Mark A.\ van de Wiel]{
Dept.\ of Epidemiology \& Biostatistics \\
VU University medical center Amsterdam \\
Amsterdam\\
The Netherlands; \and
Dept.\ of Mathematics \\
VU University Amsterdam \\
Amsterdam\\
The Netherlands}
\email{mark.vdwiel@vumc.nl}

\author[W.N.\ van Wieringen]{Wessel N.\ van Wieringen}
\address[Wessel N.\ van Wieringen]{
Dept.\ of Epidemiology \& Biostatistics \\
VU University medical center Amsterdam \\
Amsterdam\\
The Netherlands; \and
Dept.\ of Mathematics \\
VU University Amsterdam \\
Amsterdam\\
The Netherlands}
\email{w.vanwieringen@vumc.nl}

\begin{abstract}\label{abstract}
\noindent Many modern statistical applications ask for the
estimation of a covariance (or precision) matrix in settings where the number of
variables is larger than the number of observations. There exists a
broad class of ridge-type estimators that employs regularization to
cope with the subsequent singularity of the sample covariance
matrix. These estimators depend on a penalty parameter and choosing
its value can be hard, in terms of being computationally unfeasible
or tenable only for a restricted set of ridge-type estimators. Here
we introduce a simple graphical tool, the spectral condition number
plot, for informed heuristic penalty parameter selection. The
proposed tool is computationally friendly and can be employed for
the full class of ridge-type covariance (precision) estimators.

\begin{sloppypar}
\end{sloppypar}
\end{abstract}

\maketitle

\section{Introduction}\label{Intro}
The covariance matrix $\mathbf{\Sigma}$ of a $p$-dimensional random
vector $Y_i^{\mathrm{T}} \equiv [Y_{i1}, \ldots, Y_{ip}] \in
\mathbb{R}^{p}$ is of central importance in many statistical
analysis procedures. Estimation of $\mathbf{\Sigma}$ or its inverse
$\mathbf{\Sigma}^{-1} \equiv \mathbf{\Omega}$ (generally known as
the precision matrix) are central to, for example, multivariate
regression, time-series analysis, canonical correlation analysis,
discriminant analysis, and Gaussian graphical modeling. Let
$\boldsymbol{\mathrm{y}}_{i}^{\mathrm{T}}$ be a realization of
$Y_i^{\mathrm{T}}$. It is well-known \citep{Stein75} that the sample
covariance matrix $\hat{\mathbf{\Sigma}} =
n^{-1}\sum_{i=1}^{n}(\boldsymbol{\mathrm{y}}_{i} -
\bar{\boldsymbol{\mathrm{y}}})(\boldsymbol{\mathrm{y}}_{i} -
\bar{\boldsymbol{\mathrm{y}}})^{\mathrm{T}}$ is a poor estimate of
$\mathbf{\Sigma}$ when $p$ approaches the sample size $n$ or when $p
> n$. When $p$ approaches $n$, $\hat{\mathbf{\Sigma}}$ will tend to
become ill-conditioned. When $p
> n$, $\hat{\mathbf{\Sigma}}$ is singular leaving
$\hat{\mathbf{\Omega}}$ undefined. However, many contemporary
applications in fields such as molecular biology, neuroimaging, and
finance, encounter situations of interest where $p > n$.

The estimation of $\mathbf{\Sigma}$ can be improved by shrinking the
eigenvalues of $\hat{\mathbf{\Sigma}}$ towards a central value
\citep[e.g.,][]{Stein75} or by convexly combining
$\hat{\mathbf{\Sigma}}$ with some well-conditioned target matrix
\citep[e.g.,][]{SS05}. These solutions define a class of estimators
that can be caught under the umbrella term `ridge-type covariance
estimation'. Such estimators depend on a penalty parameter
determining the rate of shrinkage and choosing its value is of prime
importance. Available procedures for choosing the penalty have some
(situation dependent) disadvantages in the sense that (a) they can
be computationally expensive, (b) they can be restricted to special
cases within the class of ridge-type estimators, or (c) they are not
guaranteed to result in a meaningful penalty-value. There is thus
some demand for a generic and computationally friendly procedure on
the basis of which one can (i) heuristically select an acceptable
(minimal) value for the penalty parameter and (ii) assess if more
formal procedures have indeed proposed an acceptable penalty-value.
Here such a tool is provided on the basis of the matrix condition
number.

The remainder is organized as follows. Section \ref{Eigen} reviews
the class of ridge-type estimators of the covariance matrix. In
addition, penalty parameter selection is reviewed and an expos\'{e}
of the matrix condition number is given. The spectral condition
number is central to the introduction of the spectral condition
number plot in Section \ref{CNplot}. This graphical display is
posited as an exploratory tool, that may function as a fast and
simple heuristic in determining the (minimum) value of the penalty
parameter when employing ridge estimators of the covariance or
precision matrix. Section \ref{Illustrate} illustrates usage of the
spectral condition number plot with data from the field of
oncogenomics. Section \ref{Software} discourses on the software
that implements the proposed graphical display. Sections
\ref{Discuss} and \ref{Conclude} conclude with discussions.

\section{Eigenstructure regularization and the condition number}\label{Eigen}
\subsection{Ridge-type shrinkage estimation}\label{Ridges}
Regularization of the covariance matrix goes back to Stein
\citep{Stein75, Stein86}, who proposed shrinking the sample
eigenvalues towards some central value. This work spurred a large
body of literature \citep[see, e.g.,][]{Haff80, Haff91, YB94, Won2013}.
Of late, two encompassing forms of what is referred to as
`ridge-type covariance estimation' have emerged.

The first form considers convex combinations of
$\hat{\mathbf{\Sigma}}$ and some positive definite (p.d.) target
matrix $\mathbf{T}$
\citep{DGK,Ledo2003,LW_port_04,Ledo2004,SS05,Fish11}:
\begin{equation}\label{Steinridge}
\hat{\mathbf{\Sigma}}^{\mathrm{I}}(\lambda_{\mathrm{I}}) =
(1-\lambda_{\mathrm{I}}) \hat{\mathbf{\Sigma}} + \lambda_{\mathrm{I}}
\mathbf{T},
\end{equation}
with penalty parameter $\lambda_{\mathrm{I}} \in (0,1]$. Such an
estimator can be motivated from the Steinian bias-variance tradeoff
as it seeks to balance the high-variance, low-bias matrix
$\hat{\mathbf{\Sigma}}$ with the lower-variance, higher-bias matrix
$\mathbf{T}$. The second form is of more recent vintage and
considers the ad-hoc estimator \citep{Wart08, YC08, HA14}:
\begin{equation}\label{Adhocridge}
\hat{\mathbf{\Sigma}}^{\mathrm{II}}(\lambda_{\mathrm{II}}) =
\hat{\mathbf{\Sigma}} + \lambda_{\mathrm{II}} \mathbf{I}_{p},
\end{equation}
with $\lambda_{\mathrm{II}} \in (0, \infty)$. This second form is
motivated, much like how ridge regression was introduced by
\citet{Hoer1970}, as an ad-hoc modification of
$\hat{\mathbf{\Sigma}}$ in order to deal with singularity in the
high-dimensional setting.

\citet{GraphRidge} show that both (\ref{Steinridge}) and
(\ref{Adhocridge}) can be justified as penalized maximum likelihood
estimators \citep[cf.][]{Wart08}. However, neither (\ref{Steinridge})
nor (\ref{Adhocridge}) utilizes a proper $\ell_2$-penalty in that
perspective. Starting from the proper ridge-type $\ell_2$-penalty
$\frac{\lambda_{a}}{2}\mbox{tr}\left[(\mathbf{\Omega} -
\mathbf{T})^{\mathrm{T}}(\mathbf{\Omega} -
\mathbf{T})\right]$, \citet{GraphRidge} derive an alternative
estimator:
\begin{equation}\label{Properridge}
\hat{\mathbf{\Sigma}}^{a}(\lambda_{a}) =
\left[\lambda_{a}\mathbf{I}_{p} + \frac{1}{4}(\hat{\mathbf{\Sigma}} -
\lambda_{a}\mathbf{T})^{2}\right]^{1/2} + \frac{1}{2}(\hat{\mathbf{\Sigma}} -
\lambda_{a}\mathbf{T}),
\end{equation}
with $\lambda_{a} \in (0, \infty)$. \citet{GraphRidge} show that,
when considering a p.d.\ $\mathbf{T}$, the estimator
(\ref{Properridge}) is an alternative to (\ref{Steinridge}) with
superior behavior in terms of risk. When considering the non-p.d.\
choice $\mathbf{T} = \boldsymbol{0}$, they show that
(\ref{Properridge}) acts as an alternative to (\ref{Adhocridge}),
again with superior behavior.

Clearly, one may obtain ridge estimators of the precision matrix by
considering the inverses of (\ref{Steinridge}), (\ref{Adhocridge}),
and (\ref{Properridge}). For comparisons of these estimators see
\citep{LP85,DK01,GraphRidge}. For expositions of other penalized
covariance and precision estimators we confine by referring to
\citep{Pour13}.

\subsection{Penalty parameter selection}\label{PenaltySelect}
The choice of penalty-value is crucial to the aforementioned
estimators. Let $\lambda$ denote a generic penalty. When choosing
$\lambda$ too small, an ill-conditioned estimate may ensue when $p >
n$ (see Section \ref{Connumber}). When choosing $\lambda$ too large,
relevant data signal may be lost. Many options for choosing
$\lambda$ are available. The ridge estimators, in contrast to
$\ell_1$-regularized estimators of the covariance or precision
matrix \citep[e.g.,][]{Frie2008, SparseCov}, do not generally produce
sparse estimates. This implies that model-selection-consistent
methods (such as usage of the BIC), are not appropriate. Rather, for
$\ell_2$-type estimators, it is more appropriate to seek loss
efficiency.

A generic strategy for determining an optimal value for $\lambda$
that can be used for any ridge-type estimator of Section
\ref{Ridges} is $k$-fold cross-validation (of the likelihood
function). Asymptotically, such an approach can be explained in
terms of minimizing Kullback-Leibler divergence. Unfortunately, this
strategy is computationally prohibitive for large $p$ and/or large
$n$. \citet{Lian2011} and \citet{Vuja2014} propose approximate
solutions to the leave-one-out cross-validation score. While these
approximations imply gains in computational efficiency, they are not
guaranteed to propose a reasonable optimal value for $\lambda$.

\citet{Ledo2004} propose a strategy to determine analytically an
optimal value for $\lambda$ under a modified Frobenius loss for the
estimator (\ref{Steinridge}) under certain choices of $\mathbf{T}$
\citep[cf.][]{Fish11}. This optimal value requires information on the
unknown population matrix $\mathbf{\Sigma}$. The optimal penalty-value thus needs to be approximated with some estimate of
$\mathbf{\Sigma}$. \citet{Ledo2004} utilize an $n$-consistent
estimator while \citet{SS05} use the unbiased estimate
$\tfrac{n}{n-1}\hat{\mathbf{\Sigma}}$. In practice, this may result in
overshrinkage \citep{DK01} or even negative penalty-values
\citep{SS05}.

Given the concerns stated above, there is some demand for a generic
and computationally friendly tool for usage in the following
situations of interest:
\begin{enumerate}[i.]
  \item When one wants to speedily determine a (minimal)
value for $\lambda$ for which $\hat{\mathbf{\Sigma}}(\lambda)$ is
well-conditioned;
  \item When one wants to determine speedily whether an optimal
  $\lambda$ proposed by some other (formal) procedure indeed
  leads to a well-conditioned estimate $\hat{\mathbf{\Sigma}}(\lambda)$;
  \item When one wants to determine speedily a reasonable
  minimal value for $\lambda$ for usage in a search-grid (for
  an optimal such value) by other, optimization-based, procedures.
\end{enumerate}
In Section \ref{CNplot} we propose such a tool based on the spectral
condition number.

\subsection{Spectral condition number}\label{Connumber}
The estimators from Section \ref{Ridges} are p.d.\ when their penalty-values are strictly positive. However, they are not necessarily
well-conditioned for any strictly positive penalty-value when $p
\gtrapprox n$, especially when the penalty takes a value in the
lower range. We seek to quantify the condition of the estimators
w.r.t.\ a given penalty-value. To this end we utilize a condition
number \citep{Neu47,Turing48}.

Consider a nonsingular matrix $\mathbf{A} \in \mathbb{R}^{p \times
p}$ as well as the matrix norm $\|\cdot\|$. The condition number
w.r.t.\ matrix inversion can be defined as \citep{High95}
\begin{equation}\label{CondNumber}
  \mbox{cond}(\mathbf{A}) := \lim_{\epsilon\rightarrow 0^{+}} \sup_{\| \delta\mathbf{A} \| \leq \epsilon\| \mathbf{A} \|}
  \frac{\| (\mathbf{A} + \delta\mathbf{A})^{-1} - \mathbf{A}^{-1} \|}{\epsilon\| \mathbf{A}^{-1} \|},
\end{equation}
indicating the sensitivity of inversion of $\mathbf{A}$ w.r.t.\ small
perturbations $\delta\mathbf{A}$. When the norm in
(\ref{CondNumber}) is induced by a vector norm the condition number
is characterized as \citep{High95}:
\begin{equation}\label{IndCondNumber}
  \mbox{cond}(\mathbf{A}) = \mathcal{C}(\mathbf{A}) := \| \mathbf{A} \| \| \mathbf{A}^{-1} \|.
\end{equation}
For singular matrices $\mathcal{C}(\mathbf{A})$ would equal
$\infty$. Hence, a high condition number is indicative of
near-singularity and quantifies an ill-conditioned matrix. Indeed,
the inverse of the condition number gives the relative distance of
$\mathbf{A}$ to the set of singular matrices $\mathcal{S}$
\citep{Demmel87}:
\begin{equation*}\label{Dist}
  \mbox{dist}(\mathbf{A}, \mathcal{S}) = \inf_{\mathbf{S} \in \mathcal{S}} \frac{\| \mathbf{A} - \mathbf{S} \|}{\| \mathbf{A} \|} = \frac{1}{\mathcal{C}(\mathbf{A})}.
\end{equation*}
Another interpretation can be found in error propagation. A high
condition number implies severe loss of accuracy or large
propagation of error when performing matrix inversion under finite
precision arithmetic. One can expect to loose at least
$\lfloor\log_{10}\mathcal{C}(\mathbf{A})\rfloor$ digits of accuracy
in computing the inverse of $\mathbf{A}$ \citep[e.g., Chapter 8 and Section 6.4 of, respectively,][]{Numeric,Gentle}.
In terms of error propagation, $\mathcal{C}(\mathbf{A})$ is also a
reasonable sensitivity measure for linear systems
$\mathbf{A}\boldsymbol{\mathrm{x}} = \boldsymbol{\mathrm{b}}$
\citep{High95} .

We can specify (\ref{IndCondNumber}) with regard to a particular
norm. We have special interest in the $\ell_2$-condition number
$\mathcal{C}_{2}(\mathbf{A})$, usually called the \emph{spectral
condition number} for its relation to the spectral decomposition.
When $\mathbf{A}$ is a symmetric p.d.\ matrix, it is well-known that
\citep{Neu47,Gentle}
\begin{equation*}\label{SCN}
  \mathcal{C}_{2}(\mathbf{A}) = \| \mathbf{A} \|_{2} \| \mathbf{A}^{-1} \|_{2} = \frac{d(\mathbf{A})_1}{d(\mathbf{A})_p},
\end{equation*}
where $d(\mathbf{A})_1 \geq \ldots \geq d(\mathbf{A})_p$ are the
eigenvalues of $\mathbf{A}$. We can connect the machinery of
ridge-type regularization to this spectral condition number.

Let $\mathbf{VD}(\hat{\mathbf{\Sigma}})\mathbf{V}^{\mathrm{T}}$ be
the spectral decomposition of $\hat{\mathbf{\Sigma}}$ with
$\mathbf{D}(\hat{\mathbf{\Sigma}})$ denoting a diagonal matrix with
the eigenvalues of $\hat{\mathbf{\Sigma}}$ on the diagonal and where
$\mathbf{V}$ denotes the matrix that contains the corresponding
eigenvectors as columns. Note that the orthogonality of $\mathbf{V}$
implies $\mathbf{VV}^{\mathrm{T}} = \mathbf{V}^{\mathrm{T}}
\mathbf{V} = \mathbf{I}_{p}$. This decomposition can then be used to
show that, like the Stein estimator \citep{Stein75}, estimate
(\ref{Adhocridge}) is rotation equivariant:
$\hat{\mathbf{\Sigma}}^{\mathrm{II}}(\lambda_{\mathrm{II}}) =
\mathbf{VD}(\hat{\mathbf{\Sigma}})\mathbf{V}^{\mathrm{T}} +
\lambda_{\mathrm{II}} \mathbf{VV}^{\mathrm{T}} =
\mathbf{V}[\mathbf{D}(\hat{\mathbf{\Sigma}}) +
\lambda_{\mathrm{II}}\mathbf{I}_{p}]\mathbf{V}^{\mathrm{T}}$. That
is, the estimator leaves the eigenvectors of $\hat{\mathbf{\Sigma}}$
intact and thus solely performs shrinkage on the eigenvalues. When
choosing $\mathbf{T} = \varphi\mathbf{I}_{p}$ with $\varphi \in [0,
\infty)$, the estimator (\ref{Properridge}) also is of the rotation
equivariant form, as we may then write:
\begin{equation}\label{ProperDecomp}
\hat{\mathbf{\Sigma}}^{a}(\lambda_{a}) =
\mathbf{V}\left\{\left[\lambda_{a}\mathbf{I}_{p} +
\frac{1}{4}[\mathbf{D}(\hat{\mathbf{\Sigma}}) -
\lambda_{a}\varphi\mathbf{I}_{p}]^{2}\right]^{1/2} +
\frac{1}{2}[\mathbf{D}(\hat{\mathbf{\Sigma}}) -
\lambda_{a}\varphi\mathbf{I}_{p}]\right\}\mathbf{V}^{\mathrm{T}}.
\end{equation}
An analogous decomposition can be given for the estimator
(\ref{Steinridge}) when choosing $\mathbf{T} = \mu \mathbf{I}_{p}$,
with $\mu \in (0, \infty)$. The effect of the shrinkage estimators
can then, using the rotation equivariance property, also be
explained in terms of restricting the spectral condition number. For
example, using (\ref{ProperDecomp}), we have:
\begin{equation*}\label{Workings}
1 \leq \frac{\sqrt{\lambda_a + [d(\hat{\mathbf{\Sigma}})_{1} -
\lambda_a\varphi]^{2}/4} + [d(\hat{\mathbf{\Sigma}})_{1} -
\lambda_a\varphi]/2} {\sqrt{\lambda_a +
[d(\hat{\mathbf{\Sigma}})_{p} - \lambda_a\varphi]^{2}/4} +
[d(\hat{\mathbf{\Sigma}})_{p} - \lambda_a\varphi]/2} <
\frac{d(\hat{\mathbf{\Sigma}})_1}{d(\hat{\mathbf{\Sigma}})_p} \leq
\infty.
\end{equation*}
Similar statements can be made regarding all rotation equivariant
versions of the estimators discussed in Section \ref{Ridges}.
Similar statements can also be made when considering a target
$\mathbf{T}$ for estimators (\ref{Steinridge}) and
(\ref{Properridge}) that is not of the form $\alpha\mathbf{I}_{p}$
whenever this target is well-conditioned (i.e., has a lower
condition number than $\hat{\mathbf{\Sigma}}$).

\begin{Example}
For clarification consider the following toy example.
Assume we have a sample covariance matrix $\hat{\mathbf{\Sigma}}$ whose largest eigenvalue $d(\hat{\mathbf{\Sigma}})_1 = 3$.
Additionally assume that $\hat{\mathbf{\Sigma}}$ is estimated in a situation where $p > n$ so that $d(\hat{\mathbf{\Sigma}})_p = 0$ and, hence,
$d(\hat{\mathbf{\Sigma}})_1/d(\hat{\mathbf{\Sigma}})_p = 3/0 = \infty$ (under the IEEE computing Standard for Floating-Point Arithmetic; \citep{IEEEstandard}).
Say we are interested in regularization using the estimator (\ref{Properridge}) using a scalar target matrix with $\varphi = 2$.
Even using a very small penalty of $\lambda_{a} = \num{1e-10}$ it is then quickly verified that
$d[\hat{\mathbf{\Sigma}}^{a}(\lambda_{a})]_1/d[\hat{\mathbf{\Sigma}}^{a}(\lambda_{a})]_p = 300,003 < d(\hat{\mathbf{\Sigma}})_1/d(\hat{\mathbf{\Sigma}})_p$.
Under a large penalty such as $\lambda_{a} = 10,000$ we find that $d[\hat{\mathbf{\Sigma}}^{a}(\lambda_{a})]_1/d[\hat{\mathbf{\Sigma}}^{a}(\lambda_{a})]_p = 1.00015$.
Indeed, \citet{GraphRidge} have shown that, in this rotation equivariant setting, $d[\hat{\mathbf{\Sigma}}^{a}(\lambda_{a})]_j \rightarrow 1/\varphi$ as $\lambda_{a} \rightarrow \infty$ for all $j$.
Hence, $d[\hat{\mathbf{\Sigma}}^{a}(\lambda_{a})]_1/d[\hat{\mathbf{\Sigma}}^{a}(\lambda_{a})]_p \rightarrow  1$ as the penalty value grows very large.
Section 1 of the Supplementary Material visualizes this behavior (in the given setting) for various scenarios of interest:
$\varphi < 1/d(\hat{\mathbf{\Sigma}})_1$, $1/d(\hat{\mathbf{\Sigma}})_1 < \varphi < 1$, $\varphi = 1$, $1 < \varphi < d(\hat{\mathbf{\Sigma}})_1$, and $\varphi > d(\hat{\mathbf{\Sigma}})_1$.
\QEDE
\end{Example}

Let $\hat{\mathbf{\Sigma}}(\lambda)$ denote a generic ridge-type
estimator of the covariance matrix under generic penalty $\lambda$.
We thus quantify the conditioning of
$\hat{\mathbf{\Sigma}}(\lambda)$ for given $\lambda$ (and possibly a
given $\mathbf{T}$) w.r.t.\ perturbations $\lambda + \delta\lambda$
with
\begin{equation}\label{SCNridge}
  \mathcal{C}_{2}[\hat{\mathbf{\Sigma}}(\lambda)] = \| \hat{\mathbf{\Sigma}}(\lambda) \|_{2} \| \hat{\mathbf{\Omega}}(\lambda) \|_{2} =
  \frac{d[\hat{\mathbf{\Sigma}}(\lambda)]_1}{d[\hat{\mathbf{\Sigma}}(\lambda)]_p}.
\end{equation}
A useful property is that
$\mathcal{C}_{2}[\hat{\mathbf{\Sigma}}(\lambda)] =
\mathcal{C}_{2}[\hat{\mathbf{\Omega}}(\lambda)]$, i.e., knowing the
condition of the covariance matrix implies knowing the condition of
the precision matrix (so essential in contemporary topics such as
graphical modeling). The condition number (\ref{SCNridge}) can be
used to construct a simple and computationally friendly visual tool
for penalty parameter selection.

\section{The spectral condition number plot}\label{CNplot}
\subsection{The basic plot}\label{Basic}
As one may appreciate from the exposition in Section
\ref{Connumber}, when $\hat{\mathbf{\Sigma}}(\lambda)$ moves away
from near-singularity, small increments in the penalty $\lambda$ can
cause dramatic changes in
$\mathcal{C}_{2}[\hat{\mathbf{\Sigma}}(\lambda)]$. One can expect
that at some point along the domain of $\lambda$, its value will be
large enough for $\mathcal{C}_{2}[\hat{\mathbf{\Sigma}}(\lambda)]$
to stabilize (in some relative sense). We will cast these
expectations regarding the behavior in a (loose) definition:

\begin{heuristic*}\label{Conditioned}
Let $\hat{\mathbf{\Sigma}}(\lambda)$ denote a generic ridge-type
estimator of the covariance matrix under generic fixed penalty
$\lambda$. In addition, let $\Delta\lambda$ indicate a real
perturbation in $\lambda$ as opposed to the theoretical perturbation
$\delta\lambda$. We will term the estimate
$\hat{\mathbf{\Sigma}}(\lambda)$ \emph{well-conditioned} when small
increments $\Delta\lambda$ in $\lambda$ translate to (relatively)
small changes in $\mathcal{C}_{2}[\hat{\mathbf{\Sigma}}(\lambda +
\Delta\lambda)]$ vis-\`{a}-vis
$\mathcal{C}_{2}[\hat{\mathbf{\Sigma}}(\lambda)]$.
\end{heuristic*}

\begin{figure}[t!]
  \includegraphics[scale = .27]{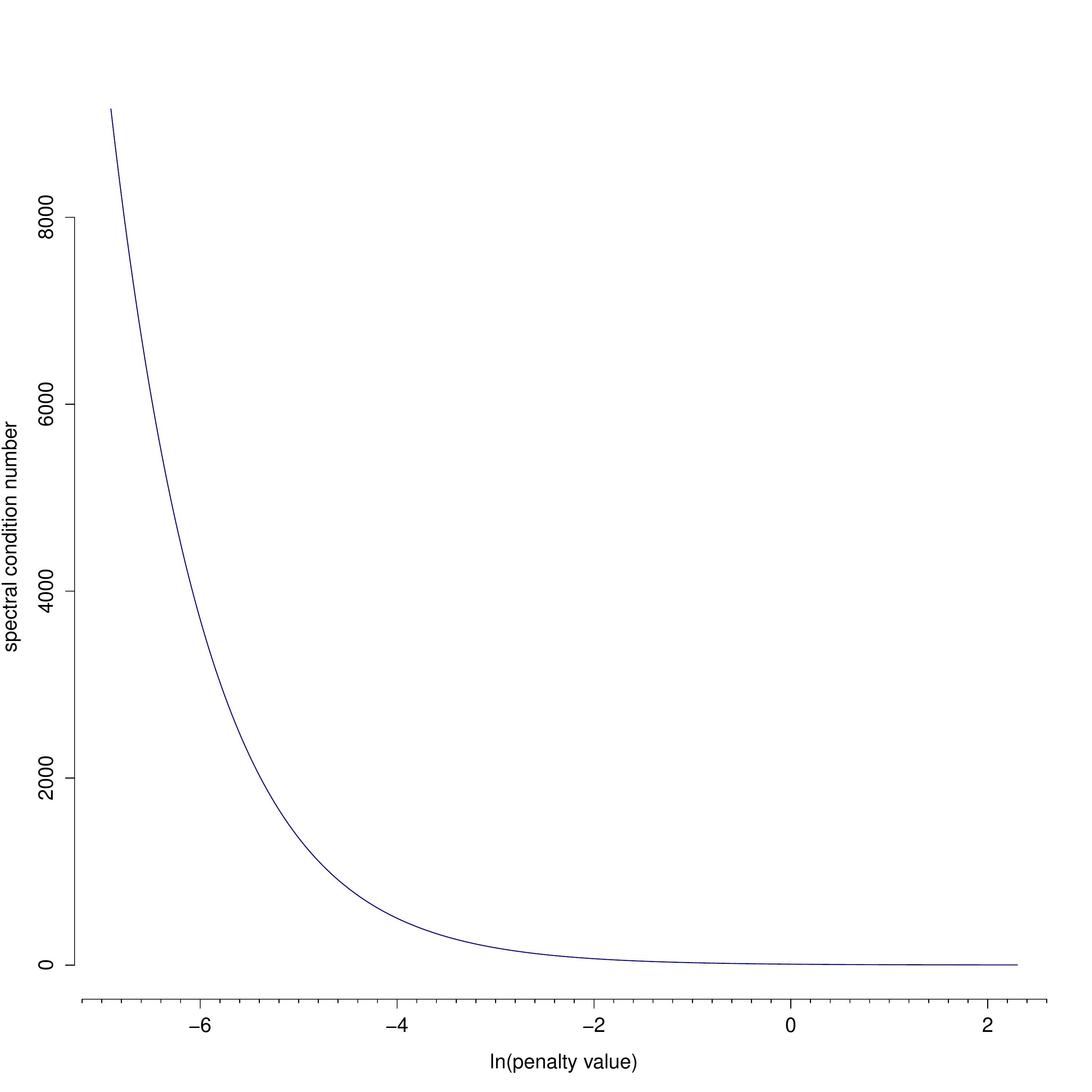}\\
  \caption{First example of a spectral condition number plot. The ridge estimator used
  is given in (\ref{Adhocridge}). \textsf{R} code to generate the data and produce the plot can be found in
  Section 4 of the Supplementary Material. A reasonable minimal value for the penalty parameter can be found at
  the $x$-axis at approximately $-3$. The exponent of this number signifies a minimal value of the penalty for which the
  estimate $\hat{\mathbf{\Sigma}}(\lambda)$ is well-conditioned according to the Heuristic
  Definition. The condition number $\mathcal{C}_{2}[\hat{\mathbf{\Sigma}}^{\mathrm{II}}(\exp(-3))] \approx 184.95$.}\label{SCNplot}
\end{figure}

From experience, when considering ridge-type estimation of
$\mathbf{\Sigma}$ or its inverse in $p > n$ situations, the point of
relative stabilization can be characterized by a leveling-off of the
acceleration along the curve when plotting
the condition number
$\mathcal{C}_{2}[\hat{\mathbf{\Sigma}}(\lambda)]$ against the (chosen) domain
of $\lambda$. Consider Figure \ref{SCNplot}, which is the first
example of what we call the \emph{spectral condition number plot}.

Figure \ref{SCNplot} indeed plots (\ref{SCNridge}) against the
natural logarithm of $\lambda$. As should be clear from Section \ref{Connumber},
the spectral condition number displays (mostly) decreasing concave upward behavior
in the feasible domain of the penalty parameter with a vertical asymptote at $\ln(0)$ and a
horizontal asymptote at $\mathcal{C}_{2}(\mathbf{T})$ (which amounts to $1$ in case of a strictly positive scalar target).
The logarithm is used on the $x$-axis
as, especially for estimators (\ref{Adhocridge}) and
(\ref{Properridge}), it is more natural to consider orders of
magnitude for $\lambda$. In addition, usage of the logarithm
`decompresses' the lower domain of $\lambda$, which enhances the
visualization of the point of relative stabilization, as it is in the lower domain of the penalty
parameter where ill-conditioning usually ensues when $p > n$. Figure
\ref{SCNplot} uses simulated data (see Section 4 of the Supplementary Material) with $p
= 100$ and $n = 25$. The estimator used is the ad-hoc ridge-type
estimator given by (\ref{Adhocridge}). One can observe relative
stabilization of the spectral condition number -- in the sense of
the Heuristic Definition -- at approximately $\exp(-3)\approx
.05$. This value can be taken as a reasonable
(minimal) value for the penalty parameter. The spectral condition
number plot can be a simple visual tool of interest in the
situations sketched in Section \ref{PenaltySelect}, as will be illustrated in Section \ref{Illustrate}.

\subsection{Interpretational aids}\label{IAid}
The basic spectral condition number plot can be amended with interpretational aids.
The software (see Section \ref{Software}) can add two such aids to form a panel of plots.
These aids support the heuristic choice for a penalty-value and provide additional information on the basic plot.

\begin{figure}
\centering
  \includegraphics[scale = .39]{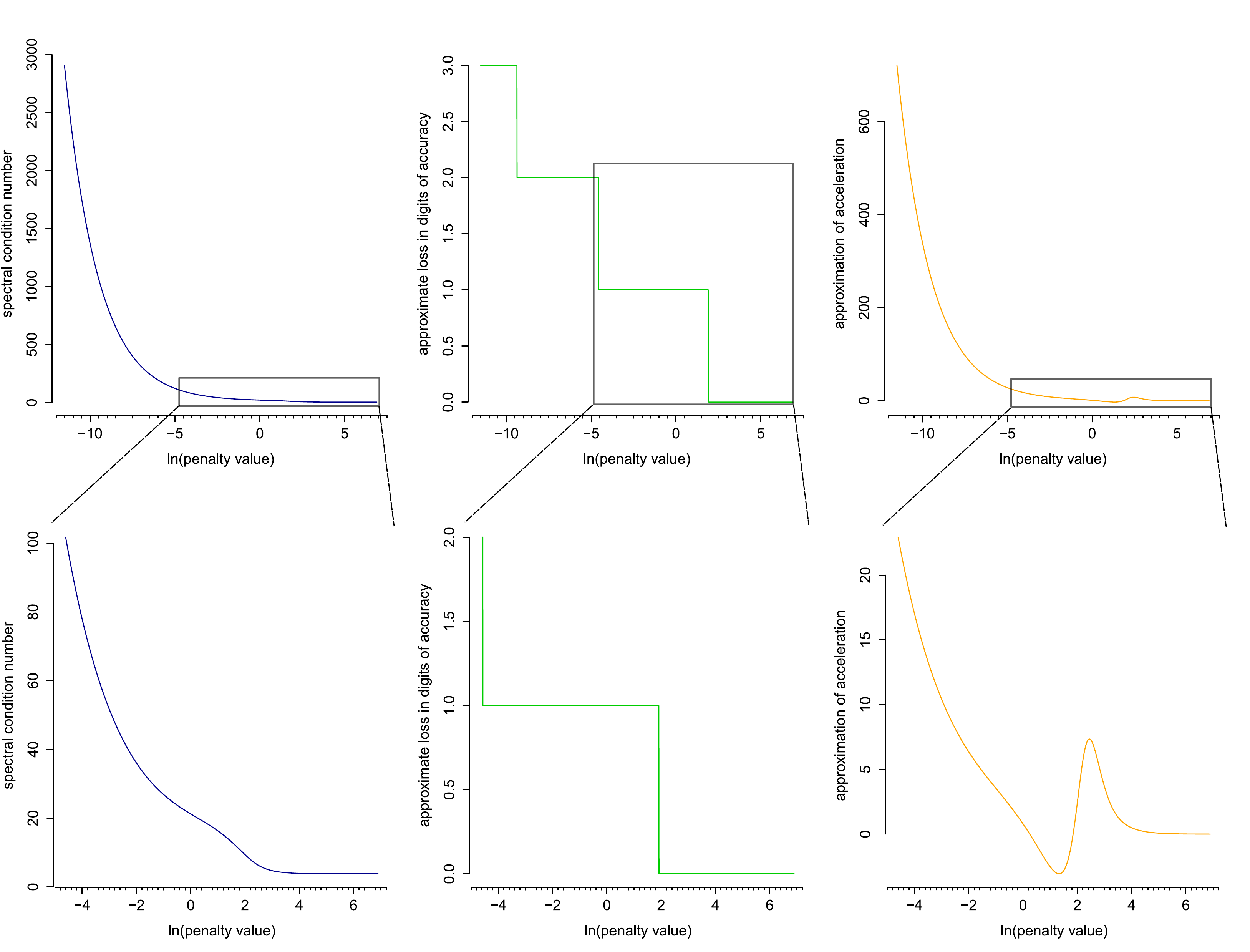}
    \caption{The spectral condition number plot with interpretational aids.
The data are the same as for Figure \ref{SCNplot} (see Section 4 of the Supplementary Material).
The ridge estimator used is given in (\ref{Properridge}) with target $\mathbf{T} = (\hat{\mathbf{\Sigma}} \circ \mathbf{I}_p)^{-1}$.
This estimator exhibits nonlinear shrinkage.
The left-hand panels give the basic spectral condition number plot.
The middle and right-hand panels exemplify the interpretational aids to the basic plot:
the approximate loss in digits of accuracy (middle panel) and the approximation of the acceleration along the curve in the basic plot (right-hand panel).
The top panels give the basic condition number plot and its interpretational aids for the domain $\lambda_{a} \in [\num{1e-5},1000]$.
The bottom panels zoom in on the boxed areas.
The interpretational aids can support the selection of a (minimal) penalty-value and may provide additional information on the basic plot.
For example, say we are interested in choosing (approximately) the minimal value for the penalty for which the error propagation (in terms of approximate loss in digits of accuracy) is at most 1.
From the middle panels we see that we should then choose the penalty-value to be $\exp(-4.2)$.
From the right-hand panels we may infer that the regularization path of the condition number displays decreasing concave downward behavior for penalty-values between approximately $\exp(.2)$ and $\exp(1.8)$.}
  \label{IntAids}
\end{figure}

The first aid is the visualization of $\lfloor\log_{10}\mathcal{C}_{2}[\hat{\mathbf{\Sigma}}(\lambda)]\rfloor$ against the domain of $\lambda$.
As stated in Section \ref{Connumber}, $\lfloor\log_{10}\mathcal{C}_{2}[\hat{\mathbf{\Sigma}}(\lambda)]\rfloor$ provides an estimate of the digits of accuracy one can expect to loose (on top of the digit loss due to inherent numerical imprecision) in operations based on $\hat{\mathbf{\Sigma}}(\lambda)$.
Note that this estimate is dependent on the norm.
This aid can support choosing a (miminal) penalty-value on the basis of the error propagation (in terms of approximate loss in digits of accuracy) one finds acceptable.
Figure \ref{IntAids} gives an example.

Let $\mathfrak{C}_{\ln(\lambda) \mapsto \mathcal{C}_{2}[\hat{\mathbf{\Sigma}}(\lambda)]}$ denote the curvature (of the basic plot) that maps the natural logarithm of the penalty-value to the condition number of the regularized precision matrix.
We seek to approximate the second-order derivative of this curvature (the acceleration) at given penalty-values in the domain $[\lambda_{\mathrm{min}}, \lambda_{\mathrm{max}}]$.
The software (see Section \ref{Software}) requires the specification of $\lambda_{\mathrm{min}}$ and $\lambda_{\mathrm{max}}$, as well as the number of steps one wants to take along the domain $[\lambda_{\mathrm{min}}, \lambda_{\mathrm{max}}]$.
Say we take $s = 1, \ldots, S$ steps, such that $\lambda_{1} = \lambda_{\mathrm{min}}, \lambda_{2}, \ldots, \lambda_{S-1}, \lambda_{S} = \lambda_{\mathrm{max}}$.
The implementation takes steps that are log-equidistant, hence $\ln(\lambda_{s}) - \ln(\lambda_{s-1}) = [\ln(\lambda_{\mathrm{max}}) - \ln(\lambda_{\mathrm{min}})]/(S-1) \equiv \tau$ for all $s = 2,\ldots,S$.
The central finite difference approximation to the second-order derivative \citep[see e.g.,][]{FiniteDiff} of $\mathfrak{C}_{\ln(\lambda) \mapsto \mathcal{C}_{2}[\hat{\mathbf{\Sigma}}(\lambda)]}$ at $\ln(\lambda_s)$ then takes the following form:
\begin{equation}\nonumber
\mathfrak{C}^{''}_{\ln(\lambda_{s}) \mapsto \mathcal{C}_{2}[\hat{\mathbf{\Sigma}}(\lambda_{s})]} \approx
\frac{\mathcal{C}_{2}[\hat{\mathbf{\Sigma}}(\lambda_{s + 1})] - 2\mathcal{C}_{2}[\hat{\mathbf{\Sigma}}(\lambda_s)] + \mathcal{C}_{2}[\hat{\mathbf{\Sigma}}(\lambda_{s - 1})]}
{\tau^{2}},
\end{equation}
which is available for $s = 2,\ldots,S-1$.
The second visual aid thus plots $\mathfrak{C}^{''}_{\ln(\lambda) \mapsto \mathcal{C}_{2}[\hat{\mathbf{\Sigma}}(\lambda)]}$ against the feasible domain of $\lambda$ (see Figure \ref{IntAids} for an example).
The behavior of the condition number along the domain of the penalty-value is to always decrease.
This decreasing behavior is not consistently concave upward for (\ref{Steinridge}) (under general non-scalar targets) and (\ref{Properridge}), as there may be parts of the domain where the behavior is concave downward.
This aid may then more clearly indicate inflection points in the regularization path of the condition number.
In addition, this aid may put perspective on the point of relative stabilization when the $y$-axis of the basic plot represents a very wide range.

Software implementing the spectral condition number plot is
discussed in Section \ref{Software}. The following section
illustrates, using oncogenomics data, the various uses of the
spectral condition number plot with regard to covariance or
precision matrix regularization.
Section 2 of the Supplementary Material contains a second data example to further illustrate usage of the condition number plot.
Section 4 of the Supplementary Material contains all \textsf{R} code with which these illustrations can be reproduced (including querying the data).

\section{Illustration}\label{Illustrate}
\subsection{Context and data}\label{Data}
Various histological variants of kidney cancer are designated with the amalgamation `renal cell carcinoma' (RCC).
Chromophobe RCC (ChRCC) is a rather rare and predominantly sporadic histological variant, accounting for 4-6\% of RCC cases \citep{ChRCCreview}.
ChRCC originates in the distal convoluted tubule \citep{UnderstandRCC}, a portion of the nephron (the basic structural unit of the kidney) that serves to maintain electrolyte balance \citep{DCT}.
Often, histological variants of cancer have a distinct pathogenesis contingent upon the deregulation of certain molecular pathways.
A pathway can be thought of as a collection of molecular features that work interdependently to regulate some biochemical function.
Recent evidence suggests that (reactivation of) the Hedgehog (Hh) signaling pathway may support cancer development and progression in clear cell RCC (CCRCC) \citep{DormoyHedge,DamatoHedge}, the most common subtype of RCC.
The Hh-signaling pathway is crucial in the sense that it ``orchestrates tissue patterning" in embryonic development, making it ``critical to normal kidney development, as it regulates
the proliferation and differentiation of mesenchymal cells in the metanephric kidney" \citep{DamatoHedge}.
Later in life Hh-signaling is largely silenced and constitutive reactivation may elicit and support tumor growth and vascularization \citep{DormoyHedge,DamatoHedge}.
Our goal here is to explore if Hh-signaling might also be reactivated in ChRCC.
The exploration will make use of network modeling (see Section \ref{GGM}) in which the network is taken as a representation of a biochemical pathway.
This exercise hinges upon a well-conditioned precision matrix.

We attained data on RCC from the \citet{TCGAccrcc} as queried through the Cancer Genomics Data Server \citep{CeramiCbio,GaoCbio} using the \texttt{cgdsr} \textsf{R}-package \citep{CGDSR}.
All ChRCC samples were retrieved for which messenger ribonucleic acid (mRNA) data is available, giving a total of $n = 15$ samples.
The data stem from the IlluminaHiSeq\_RNASeqV2 RNA sequencing platform and consist of normalized relative gene expressions.
That is, individual gene expressions are given as mRNA z-scores relative to a reference population that consists of all tumors that are diploid for the gene in question.
All features were retained that map to the Hh-signaling pathway according to the Kyoto Encyclopedia of Genes and Genomes (KEGG) \citep{Kanehisa2000}, giving a total of $p = 56$ gene features.
Regularization of the desired precision matrix is needed as $p > n$.
Even though features from genomic data are often measured on or mapped to (approximately) the same scale, regularization on the standardized scale is often appropriate as the variability of the features may differ substantially when $p > n$; a point also made by \citep{Wart08}.
Note that we may use the correlation matrix $\mathbf{R} = (\hat{\mathbf{\Sigma}} \circ \mathbf{I}_{p})^{-\frac{1}{2}}\hat{\mathbf{\Sigma}}(\hat{\mathbf{\Sigma}} \circ \mathbf{I}_{p})^{-\frac{1}{2}}$ instead of $\hat{\mathbf{\Sigma}}$ in equations (\ref{Steinridge}) to (\ref{Properridge}) without loss of generality.

\subsection{Penalty parameter selection}\label{SelectPenalty}
The precision estimator of choice is the inverse of (\ref{Properridge}).
The target matrix is chosen as $\mathbf{T} = \varphi\mathbf{I}_{p}$, with $\varphi$ set to the reciprocal of the average eigenvalue of $\mathbf{R}$: $1$.
First, the approximate leave-one-out cross-validation (aLOOCV) procedure \citep{Lian2011,Vuja2014} is used (on the negative log-likelihood) in finding an optimal value for $\lambda_a$ under the given target and data settings.
This procedure searches for the optimal value $\lambda_{a}^{*}$ in the domain $\lambda_{a} \in [\num{1e-5},20]$.
A relatively fine-grained grid of $10,000$ log-equidistant steps along this domain points to $\num{1e-5}$ as being the optimal value for the penalty (in the chosen domain).
This value seems low given the $p/n$ ratio of the data.
This calls for usage-type ii.\ of the condition number plot (Section \ref{PenaltySelect}), where one uses it to determine if an optimal penalty as proposed by some procedure indeed leads to a well-conditioned estimate.
The condition number is plotted over the same penalty-domain considered by the aLOOCV procedure.
The left-hand panel of Figure \ref{HHnetwork} depicts this condition number plot.
The green vertical line represents the penalty-value that was chosen as optimal by the aLOOCV procedure.
Clearly, the precision estimate at $\lambda_{a} = \num{1e-5}$ is not well-conditioned in the sense of the Heuristic Definition.
This exemplifies that the (essentially large-sample) approximation to the LOOCV score may not be suitable for non-sparse situations and/or for situations in which the $p/n$ ratio grows more extreme
(the negative log-likelihood term then tends to completely dominate the bias term).
At this point one could use the condition number plot in accordance with usage-type i., in which one seeks a reasonable minimal penalty-value.
This reasonable minimal value (in accordance with the Heuristic Definition) can be found at approximately $\exp(-6)$, at which $\mathcal{C}_{2}[\hat{\mathbf{\Omega}}^{a}(\exp(-6))] = \mathcal{C}_{2}[\hat{\mathbf{\Sigma}}^{a}(\exp(-6))] \approx 247.66$.

One could worry that the precision estimate retains too much noise under the heuristic minimal penalty-value.
To this end, a proper LOOCV procedure is implemented that makes use of the root-finding Brent algorithm \citep{Brent}.
The expectation is that the proper data-driven LOOCV procedure will find an optimal penalty-value in the domain of $\lambda_{a}$ for which the estimate is well-conditioned.
The penalty-space of search is thus constrained to the region of well-conditionedness for additional speed, exemplifying usage-type iii.\ of the condition number plot.
Hence, the LOOCV procedure is told to search for the optimal value $\lambda_{a}^{*}$ in the domain $\lambda_{a} \in [\exp(-6),20]$.
The optimal penalty-value is indeed found to the right of the heuristic minimal value at $5.2$.
At this value, indicated by the red vertical line in Figure \ref{HHnetwork}, $\mathcal{C}_{2}[\hat{\mathbf{\Omega}}^{a}(5.2)] = \mathcal{C}_{2}[\hat{\mathbf{\Sigma}}^{a}(5.2)] \approx 8.76$.
The precision estimate at this penalty-value is used in further analysis.

\subsection{Further analysis}\label{GGM}
\begin{landscape}
\begin{figure}
\centering
  \includegraphics[scale = .31]{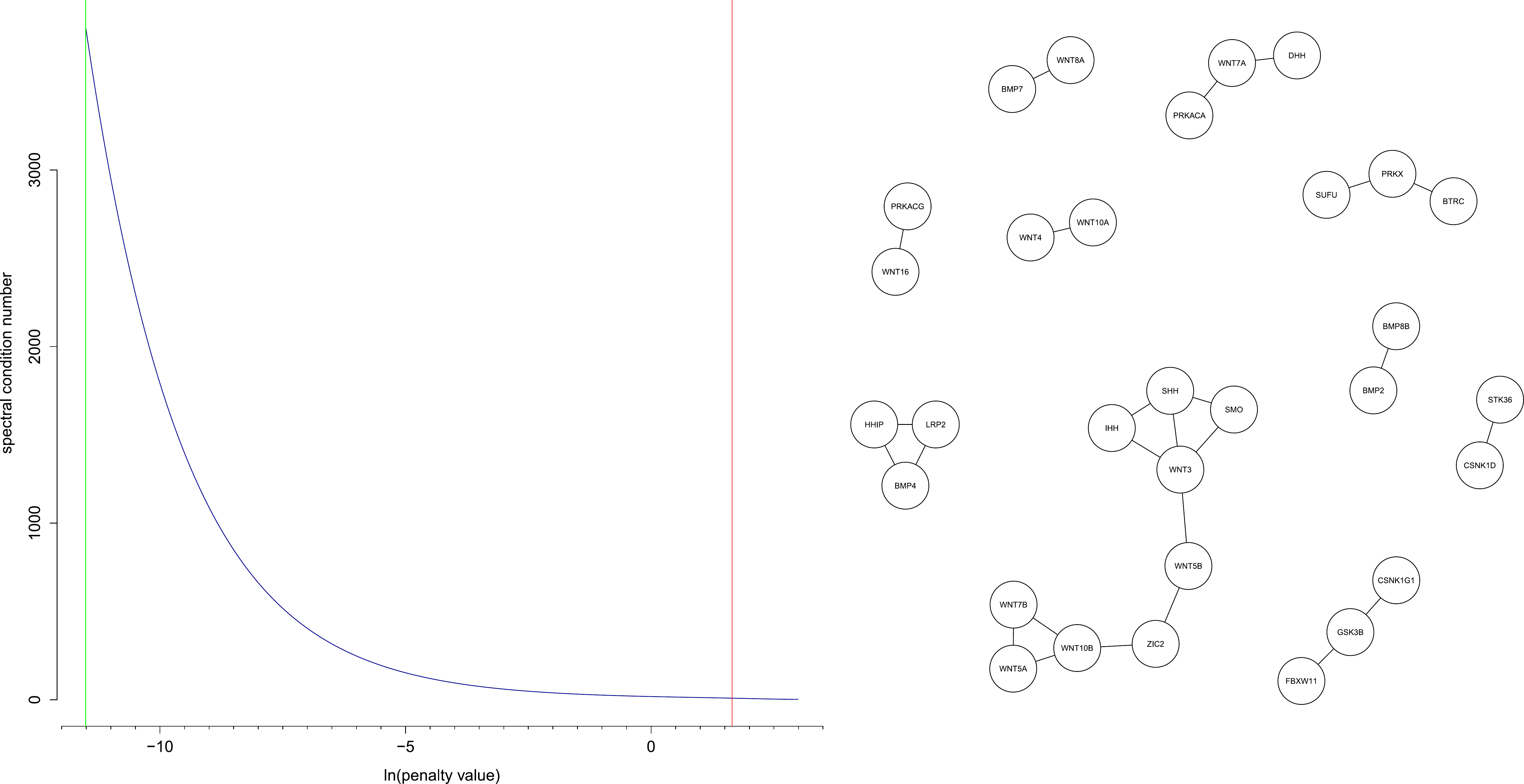}
    \caption{\emph{Left-hand panel}: Condition number plot of the Hedgehog (Hh) signaling pathway variables on the chromophobe kidney cancer data of \cite{TCGAccrcc}.
    The green vertical line indicates the value of the penalty parameter that was chosen as optimal by the aLOOCV procedure ($\num{1e-5}$).
    The red vertical line indicates the value of the penalty that was chosen as optimal by the root-finding LOOCV procedure ($5.2$).
    The value indicated by the aLOOCV procedure does not lie in a region where the estimate can be deemed well-conditioned.
    \emph{Right-hand panel}: The retrieved Markov network using the optimal penalty-value as indicated by the root-finding LOOCV procedure.
    The vertex-labels are Human Genome Organization (HUGO) Gene Nomenclature Committee (HGNC) curated gene names of the Hh-signaling pathway genes.
    Certain salient features of the Hh-signaling pathway are retrieved, indicating that this pathway may be reactivated in ChRCC.}
  \label{HHnetwork}
\end{figure}
\end{landscape}

Biochemical networks are often reconstructed from data through graphical models.
Graphical modeling refers to a class of probabilistic models that uses graphs to express conditional (in)dependence relations (i.e., Markov properties) between random variables.
Let $\mathcal{V}$ denote a finite set of vertices that correspond to a collection of random variables with probability distribution $\mathcal{P}$, i.e., $\{Y_{1},\ldots,Y_{p}\} \sim \mathcal{P}$.
Let $\mathcal{E}$ denote a set of edges, where edges are understood to consist of pairs of distinct vertices such that $Y_{j}$ is connected to $Y_{j'}$, i.e., $Y_{j} - Y_{j'} \in \mathcal{E}$.
We then consider graphs $\mathcal{G} = (\mathcal{V}, \mathcal{E})$ under the basic assumption $\{Y_{1},\ldots,Y_{p}\} \sim \mathcal{N}_{p}(\boldsymbol{0}, \mathbf{\Sigma})$.
In this Gaussian case, conditional independence between a pair of variables corresponds to zero entries in the precision matrix.
Indeed, the following relations can be shown to hold for all pairs $\{Y_{j}, Y_{j'}\} \in \mathcal{V}$ with $j \neq j'$ \citep[see, e.g.,][]{whittaker}:
\begin{equation}\nonumber
(\hat{\mathbf{\Omega}})_{jj'} = 0
\Longleftrightarrow Y_{j} \ci Y_{j'}|\mathcal{V}\setminus\{Y_{j},Y_{j'}\}
\Longleftrightarrow Y_{j} \centernot{-} Y_{j'},
\end{equation}
where $\centernot{-}$ indicates the absence of an edge.
Hence, the graphical model can be selected by determining the support of the precision matrix.
For support determination we resort to a local false discovery rate procedure proposed by \citep{SS05}, retaining only those edges whose empirical posterior probability of being present equals or exceeds $.80$.
The right-hand panel of Figure \ref{HHnetwork} represents the retrieved Markov network on the basis of $\hat{\mathbf{\Omega}}^{a}(5.2)$.
The vertex-labels are curated gene names of the Hh-signaling pathway genes.
The graph seems to retrieve salient features of the Hh-signaling pathway.
The Hh-signaling pathway involves a cascade from the members of the Hh-family (IHH, SHH, and DHH) via the SMO gene to ZIC2 and members of the Wnt-signaling pathway.
The largest connected component is indicative of this cascade, giving tentative evidence of reactivation of the Hh-signaling pathway in (at least) rudimentary form in ChRCC.

\section{Software}\label{Software}
\begin{sloppypar}
The \textsf{R}-package \texttt{rags2ridges} \citep{Rags} implements
(along with functionalities for graphical modeling) the ridge
estimators of Section \ref{Ridges} and the condition number plot
through the function \texttt{CNplot}. This function
outputs visualizations such as Figure \ref{SCNplot} and Figure \ref{IntAids}.
The condition number is determined by exact calculation using the spectral
decomposition. For most practical purposes this exact calculation is
fast enough, especially in rotation equivariant situations for which only
a single spectral decomposition is required to obtain the complete solution
path of the condition number. Additional computational speed is achieved by the
implementation of core operations in \textsf{C++} via \texttt{Rcpp} and \texttt{RcppArmadillo}
\cite{Eddelbuettel2011,Rcpp2013}. For example, producing the basic condition number plot
for the estimator (\ref{Properridge}) on the basis of data with dimensions $p = 1,000$ and $n = 200$,
using a scalar target matrix and $1,000$ steps along the penalty-domain, will take
approximately 1.2 seconds (see Section 3 of the Supplementary Material for a benchmark
exercise). The additional computational cost of the
interpretational aids is linear: producing the panel of plots (including interpretational aids)
for the situation just given takes approximately 1.3 seconds. When $\lambda$ is very small and $p \gg n$ the exact
calculation of the condition number may suffer from rounding problems (much like the
imaginary linear system $\mathbf{\Sigma}\boldsymbol{\mathrm{x}} =
\boldsymbol{\mathrm{b}}$), but remains adequate in its indication of
ill-conditioning.

When exact computation is deemed too costly in
terms of floating-point operations, or when one wants more speed in
determining the condition number, the \texttt{CNplot}
function offers the option to cheaply approximate the
$\ell_1$-condition number, which amounts to
\begin{equation*}
    \mathcal{C}_1[\hat{\mathbf{\Sigma}}(\lambda)] \,=\,
    \parallel\hat{\mathbf{\Sigma}}(\lambda)\parallel_{1}
    \parallel\hat{\mathbf{\Omega}}(\lambda)\parallel_{1} \,=\,
   \left\{ \max_{j'}\sum_{j}|[\hat{\mathbf{\Sigma}}(\lambda)]_{jj'}| \right\}
   \left\{ \max_{j'}\sum_{j}|[\hat{\mathbf{\Omega}}(\lambda)]_{jj'}|
   \right\}.
\end{equation*}
The $\ell_1$-condition number is computationally less complex than
the calculation of $\mathcal{C}_2[\hat{\mathbf{\Sigma}}(\lambda)]$ in non-rotation equivariant
settings. The machinery of ridge-type regularization is, however, less directly connected
to this $\ell_1$-condition number (in comparison to the $\ell_2$-condition number).
The approximation of $\mathcal{C}_1[\hat{\mathbf{\Sigma}}(\lambda)]$
uses LAPACK routines \citep{LAPACK} and avoids overflow. This
approximation is accessed through the \texttt{rcond} function from
\textsf{R} \citep{Rman}. The package \texttt{rags2ridges} is freely
available from the Comprehensive \textsf{R} Archive Network
(\url{http://cran.r-project.org/}) \citep{Rman}.
\end{sloppypar}

\section{Discussion}\label{Discuss}
The condition number plot is a heuristic tool and heuristics should be handled with care.
Below, some cases are presented that serve as notes of caution.
They exemplify that the proposed heuristic accompanying the condition number plot should not be applied (as any statistical technique) without proper inspection of the data.

\subsection{Artificial ill-conditioning}\label{Articifical}
A first concern is that, when the variables are measured on different scales, artificial ill-conditioning may ensue \citep[see, e.g.,][]{Gentle}.
In case one worries if the condition number is an adequate indication of error propagation when using variables on their original scale one can ensure that the columns (or rows) of the input matrix are on the same scale.
This is easily achieved by scaling the input covariance matrix to be the correlation matrix.
Another issue is that it is not guaranteed that the condition number plot will give an unequivocal point of relative stabilization for every data problem (which hinges in part on the chosen domain of the penalty parameter).
Such situations can be dealt with by extending the domain of the penalty parameter or by determining the value of $\lambda$ that corresponds to the loss of $\lfloor\log_{10}\mathcal{C}[\hat{\mathbf{\Sigma}}(\lambda)]\rfloor$ digits (in the imaginary linear system $\mathbf{\Sigma}\boldsymbol{\mathrm{x}} = \boldsymbol{\mathrm{b}}$) one finds acceptable.

\subsection{Naturally high condition numbers}\label{Natural}
Some covariance matrices may have high condition numbers as their `natural state'.
Consider the following covariance matrix: $\mathbf{\Sigma}_{\mbox{{\tiny equi}}} = (1- \varrho) \mathbf{I}_{p}  + \varrho \mathbf{J}_{p}$, with $-1/(p - 1) < \varrho < 1$ and where $\mathbf{J}_{p}$ denotes the $(p \times p)$-dimensional all-ones matrix.
The variates are thus equicorrelated with unit variance.
The eigenvalues of this covariance matrix are $p \varrho + (1-\varrho)$ and (with multiplicity $p-1$) $1-\varrho$.
Consequently, its condition number equals $1 + p \varrho / (1-\varrho)$.
The condition number of  $\mathbf{\Sigma}_{\mbox{{\tiny equi}}}$ thus becomes high when the number of variates grows large and/or the (marginal) correlation $\varrho$ approaches one (or $-1/(p - 1)$). The large ratio between the largest and smallest eigenvalues of $\mathbf{\Sigma}_{\mbox{{\tiny equi}}}$ in such situations mimics a high-dimensional setting in which any non-zero eigenvalue of the sample covariance estimate is infinitely larger than the smallest (zero) eigenvalues.
However, irrespective of the number of samples, any sample covariance estimate of an $\mathbf{\Sigma}_{\mbox{{\tiny equi}}}$ with large $p$ and $\varrho$ close to unity (or $-1/(p - 1)$) exhibits such a large ratio.
Would one estimate the $\mathbf{\Sigma}_{\mbox{{\tiny equi}}}$ in penalized fashion (even for reasonable  sample sizes) and choose the penalty parameter from the condition number plot as recommended, then one would select a penalty parameter that yields a `well-conditioned' estimate.
Effectively, this amounts to limiting the difference between the penalized eigenvalues, which need not give a condition number representative of $\mathbf{\Sigma}_{\mbox{{\tiny equi}}}$.
Thus, the recommendation to select the penalty parameter from the well-conditioned domain of the condition number plot may in some (perhaps exotic) cases lead to a choice that crushes too much relevant signal (shrinking the largest eigenvalue too much).
For high-dimensional settings this may be unavoidable, but for larger sample sizes this is undesirable.

\subsection{Contamination}\label{Contaminated}
Real data is often contaminated with outliers.
To illustrate the potential effect of outliers on the usage of the condition number plot, consider data $\boldsymbol{\mathrm{y}}_{i}$ drawn from a contaminated distribution, typically modeled by a mixture distribution: $\boldsymbol{\mathrm{y}}_{i} \sim (1- \phi)  \mathcal{N}_p(\mathbf{0},  \mathbf{\Sigma}) + \phi \mathcal{N}_p(\mathbf{0} , c \mathbf{I}_{p})$ for $i = 1, \dots, n$, some positive constant $c > 0$, and mixing proportion $\phi \in [0,1]$.
Then, the expectation of the sample covariance matrix $\mathbb{E}(\boldsymbol{\mathrm{y}}_{i}\boldsymbol{\mathrm{y}}_{i}^{\mathrm{T}}) = (1- \phi) \mathbf{\Sigma} + c \phi  \mathbf{I}_{p}$.
Its eigenvalues are: $d[ (1- \phi) \mathbf{\Sigma} + c \phi  \mathbf{I}_{p}]_{j}  = (1- \phi) d(\mathbf{\Sigma})_{j} + c \phi $, for $j = 1, \dots, p$.
In high-dimensional settings with few samples the presence of any outlier corresponds to mixing proportions clearly deviating from zero.
In combination with any substantial $c$ the contribution of the outlier(s) to the eigenvalues may be such that the contaminated sample covariance matrix is represented as better conditioned (vis-\`{a}-vis its uncontaminated counterpart).
It is the outlier(s) that will determine the domain of well-conditionedness in such a situation.
Then, when choosing the penalty parameter in accordance with the Heuristic Definition, undershrinkage may ensue.
One may opt, in situations in which the results are influenced by outliers, to use a robust estimator for $\hat{\mathbf{\Sigma}}$ in producing the condition number plot.

\section{Conclusion}\label{Conclude}
We have proposed a simple visual display that may be of aid in
determining the value of the penalty parameter in ridge-type
estimation of the covariance or precision matrix when the number of
variables is large relative to the sample size. The visualization we
propose plots the spectral condition number against the domain of the
penalty parameter. As the value of the penalty parameter increases,
the covariance (or precision) matrix will move away from (near)
singularity. In some lower-end of the domain this will mean that
small increments in the value of the penalty parameter will lead to
large decreases of the condition number. At some point, the
condition number can be expected to stabilize, in the sense that
small increments in the value of the penalty have (relatively)
little effect on the condition number. The point of relative
stabilization may be deemed to indicate a reasonable (minimal) value for the
penalty parameter.

Usage of the condition number plot was exemplified in situations concerned with the direct estimation of covariance or precision matrices.
The plot may also be of interest in situations in which (scaled versions of) these matrices are conducive to further computational procedures.
For example, it may support the ridge approach to the regression problem $\boldsymbol{\mathrm{x}} = \mathbf{Y}\boldsymbol{\beta} + \boldsymbol{\epsilon}$.
We would then assess the conditioning of $\mathbf{Y}^{\mathrm{T}}\mathbf{Y} + \lambda\mathbf{I}_p$
for use in the ridge-solution to the regression coefficients:
$\hat{\boldsymbol{\beta}} = (\mathbf{Y}^{\mathrm{T}}\mathbf{Y} + \lambda\mathbf{I}_p)^{-1}\mathbf{Y}^{\mathrm{T}}\boldsymbol{\mathrm{x}}$.

We explicitly state that we view the proposed condition number plot
as an \emph{heuristic tool}. We emphasize `tool', as it gives easy
and fast access to penalty-value determination without proposing an
optimal (in some sense) value. Also, in the tradition of exploratory
data analysis \citep{EDA}, usage of the condition number plot
requires good judgment. As any heuristic method, it is not without
flaws.

Notwithstanding these concerns, the condition number plot gives
access to a fast and generic (i.e., non-target and non-ridge-type
specific) procedure for regularization parameter determination that
is of use when analytic solutions are not available and when other
procedures fail. In addition, the condition number plot may aid more
formal procedures, in terms of assessing if a well-conditioned
estimate is indeed obtained, and in terms of proposing a reasonable
minimal value for the regularization parameter for usage in a search
grid.

\section*{Acknowledgements}
This research was
supported by grant FP7-269553 (EpiRadBio) through the European
Community's Seventh Framework Programme (FP7, 2007-2013).


\bibliographystyle{plainnat}
\bibliography{ConditionPlotRef}


\addresseshere

\cleardoublepage

\renewcommand{\theequation}{S\arabic{equation}}
\renewcommand{\thefigure}{S\arabic{figure}}
\renewcommand{\thetable}{S\arabic{table}}
\renewcommand{\thesection}{\arabic{section}}

\setcounter{section}{0}
\setcounter{subsection}{0}
\setcounter{equation}{0}
\setcounter{figure}{0}
\setcounter{table}{0}
\setcounter{page}{1}

\phantomsection
\addcontentsline{toc}{section}{Supplementary Material}
\begin{center}
{\huge SUPPLEMENTARY MATERIAL}
\end{center}

\bigskip
\bigskip
This supplement contains figures and \textsf{R} code in support of the main text.
In addition, this supplement contains a second illustration of (the use of) the spectral condition number plot.

\section*{1. Behavior largest and smallest eigenvalues}
\begin{figure}[ht]
  \includegraphics[scale = .3]{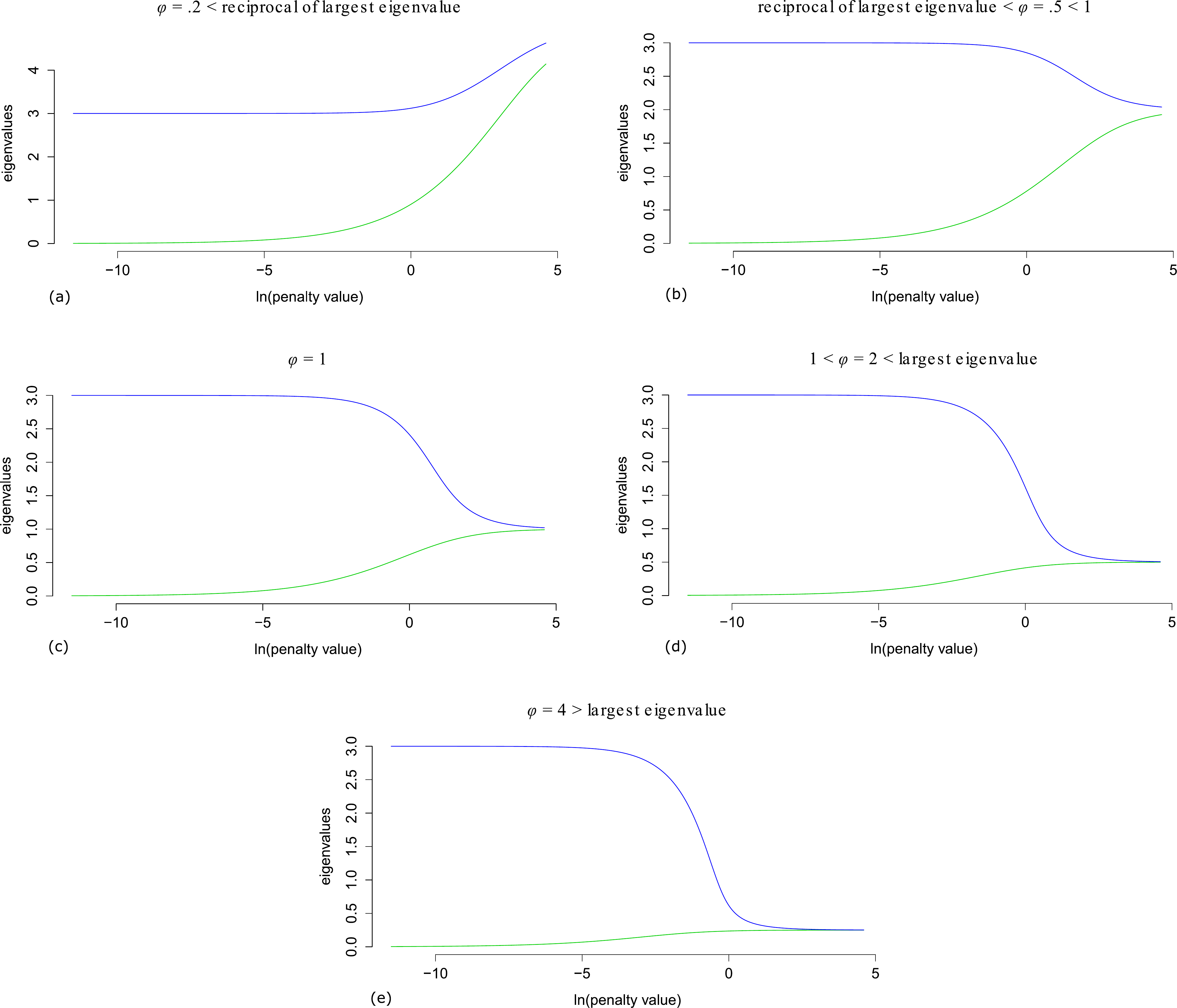}\\
  \caption{Visualization of the behavior of the largest and smallest eigenvalues along the domain of the penalty parameter.
The green line represents the smallest eigenvalue while the blue line represents the largest eigenvalue.
The panels represent different choices for the scalar $\varphi$: (a) $\varphi < 1/d(\hat{\mathbf{\Sigma}})_1$; (b) $1/d(\hat{\mathbf{\Sigma}})_1 < \varphi < 1$; (c) $\varphi = 1$; (d) $1 < \varphi < d(\hat{\mathbf{\Sigma}})_1$; and (e) $\varphi > d(\hat{\mathbf{\Sigma}})_1$.}\label{Behave}
\end{figure}

This section contain a visual impression (Figure \ref{Behave}) of the behavior of the largest and smallest eigenvalues along the domain of the penalty parameter.
The setting is given in Example 1 of the main text.

\bigskip

\newpage
\section*{2. A second Illustration}
\subsection*{2.1. Context and data}
Prostate cancer refers to an adenocarcinoma in the prostate gland.
It is the most common solid tumor diagnosed in western men \citep{CanStats16}.
Its prognosis is largely determined by metastasis with low survival rates for metastatic forms in comparison to organ-confined forms \citep{Survive}.

Vascular endothelial growth factor (VEGF) is a signal protein that supports vascularization \citep{VEGF}.
Vascular endothelial cells are cells that line the interior of blood vessels.
The formation of new blood vessels is pivotal to the growth and metastasis of solid tumors.
VEGF is actually at the helm of a cascade of signaling pathways that form the VEGF-signaling pathway.
In specific, the activation of VEGF leads to the activation of the mitogen-activated protein kinase (MAPK) and phosphatidylinositol 3'-kinase (PI3K)-AKT signaling pathways \citep{KEGGvegf}.
The VEGF-signaling pathway thus largely consists of two subpathways.
These subpathways mediate ``the proliferation and migration of endothelial cells and" promote ``their survival and vascular permeability" \citep{KEGGvegf}.
Hence, VEGF overexpression supports tumor growth and metastasis.

VEGF-signaling is likely active in metastatic prostate cancer.
This contention is supported by recent evidence that changes in the PI3K-pathway are present in metastatic tumor samples \citep{MSKCCprostate}.
Cancer may be viewed, from a pathway perspective, as consisting of a loss of normal biochemical connections and a gain of abnormal biochemical connections.
Severe deregulation would then imply the loss of normal subpathways and/or the gain of irregular subpathways.
Our goal here is to explore if the VEGF-signaling pathway in metastatic prostate cancer can still be characterized as consisting of the MAPK and PI3K-AKT subpathways.
The exploration will make use of a pathway-based factor analysis in which the retained latent factors are taken to represent biochemical subpwathways \citep[cf.][]{CarvalhoFA}.
This exercise hinges upon a well-conditioned covariance matrix.

We attained data on prostate cancer from the Memorial Sloan-Kettering Cancer Center \cite{MSKCCprostate} as queried through the Cancer Genomics Data Server \citep{CeramiCbio,GaoCbio} using the \texttt{cgdsr} \textsf{R}-package \citep{CGDSR}.
All metastatic samples were retrieved for which messenger ribonucleic acid (mRNA) data is available, giving a total of $n = 19$ samples.
The data stem from the Affymetrix Human Exon 1.0 ST array platform and consist of $\log_2$ whole-transcript mRNA expression values.
All Human Genome Organization (HUGO) Gene Nomenclature Committee (HGNC) curated features were retained that map to the VEGF-signaling pathway according to the Kyoto Encyclopedia of Genes and Genomes (KEGG) \citep{Kanehisa2000}, giving a total of $p = 75$ gene features.
Regularization of the desired covariance matrix is needed as $p > n$.
Regularization is performed (as in the illustration in the main text) on the standardized scale.

\subsection*{2.2. Model}
We work with standardized data.
Hence, let $\boldsymbol{\mathrm{z}}_i \in \mathbb{R}^{p}$ denote a centered and scaled $p$-dimensional observation vector available for $i = 1, \ldots, n$ persons.
The factor model states that
\begin{equation}\label{FACmodel}\nonumber
\boldsymbol{\mathrm{z}}_i = \mathbf{\Gamma}\boldsymbol{\mathrm{\xi}}_i + \boldsymbol{\epsilon}_i,
\end{equation}
where $\boldsymbol{\mathrm{\xi}}_i \in \mathbb{R}^{m}$ denotes an $m$-dimensional vector of latent variables typically called `factors', and where
$\mathbf{\Gamma} \in \mathbb{R}^{p \times m}$ is a matrix whose entries $\gamma_{jk}$ denote the loading of the $j$th variable on the $k$th factor,
$j = 1, \ldots, p$, $k = 1, \ldots, m$. Finally, the $\boldsymbol{\epsilon}_i \in \mathbb{R}^{p}$ denote error measurements.
Hence, the model can be conceived of as a multivariate regression with latent predictors.
An important assumption in this model is that $m < p$: the dimension of the latent vector is smaller than the dimension of the observation vector.
The following additional assumptions are made: (i) The observation vectors are independent; (ii) $\mbox{rank}(\mathbf{\Gamma}) = m$; (iii) $\boldsymbol{\mathrm{\xi}}_i \sim \mathcal{N}_m(\boldsymbol{0},\mathbf{I}_p)$; (iv) $\boldsymbol{\mathrm{\epsilon}}_i \sim \mathcal{N}_p(\boldsymbol{0},\mathbf{\Psi})$, with
$\mathbf{\Psi}$ a $(p \times p)$-dimensional diagonal matrix with strictly positive diagonal entries $\psi_j$; and (v) $\boldsymbol{\mathrm{\xi}}_i$ and $\boldsymbol{\mathrm{\epsilon}}_{i'}$ are independent for all $i$ and $i'$.
The preceding assumptions establish a distributional assumption on the covariance structure of the observed data-vector: (vi) $\boldsymbol{\mathrm{z}}_i \sim \mathcal{N}_p(\boldsymbol{0},\mathbf{\Sigma} = \mathbf{\Gamma}\mathbf{\Gamma}^{\mathrm{T}} + \mathbf{\Psi})$.

Hence, to characterize the model, we need to estimate $\mathbf{\Gamma}$ and $\mathbf{\Psi}$ on the basis of the sample correlation matrix.
As $p > n$ for the data at hand, the sample correlation matrix is singular and standard maximum likelihood estimation (MLE) is not available.
Recent efforts deal with such situations by (Bayesian) sparsity modeling of the factor model \citep[e.g.,][]{CarvalhoFA}, i.e., by imposing sparsity constraints in the loadings matrix.
Our approach differs.
We first ensure that we have a well-conditioned correlation matrix by using a regularized (essentially a Bayesian) estimator.
Afterwards, standard MLE techniques are employed to estimate $\mathbf{\Gamma}$ and $\mathbf{\Psi}$ on the basis of this regularized correlation matrix.
Hence, we perform MLE on the basis of $\hat{\mathbf{\Sigma}}(\lambda^*)$ where $\lambda^*$ is (in some sense) deemed optimal.

\subsection*{2.3. Penalty parameter selection}
\begin{figure}[tbp]
  \centering
  \includegraphics[width=\textwidth]{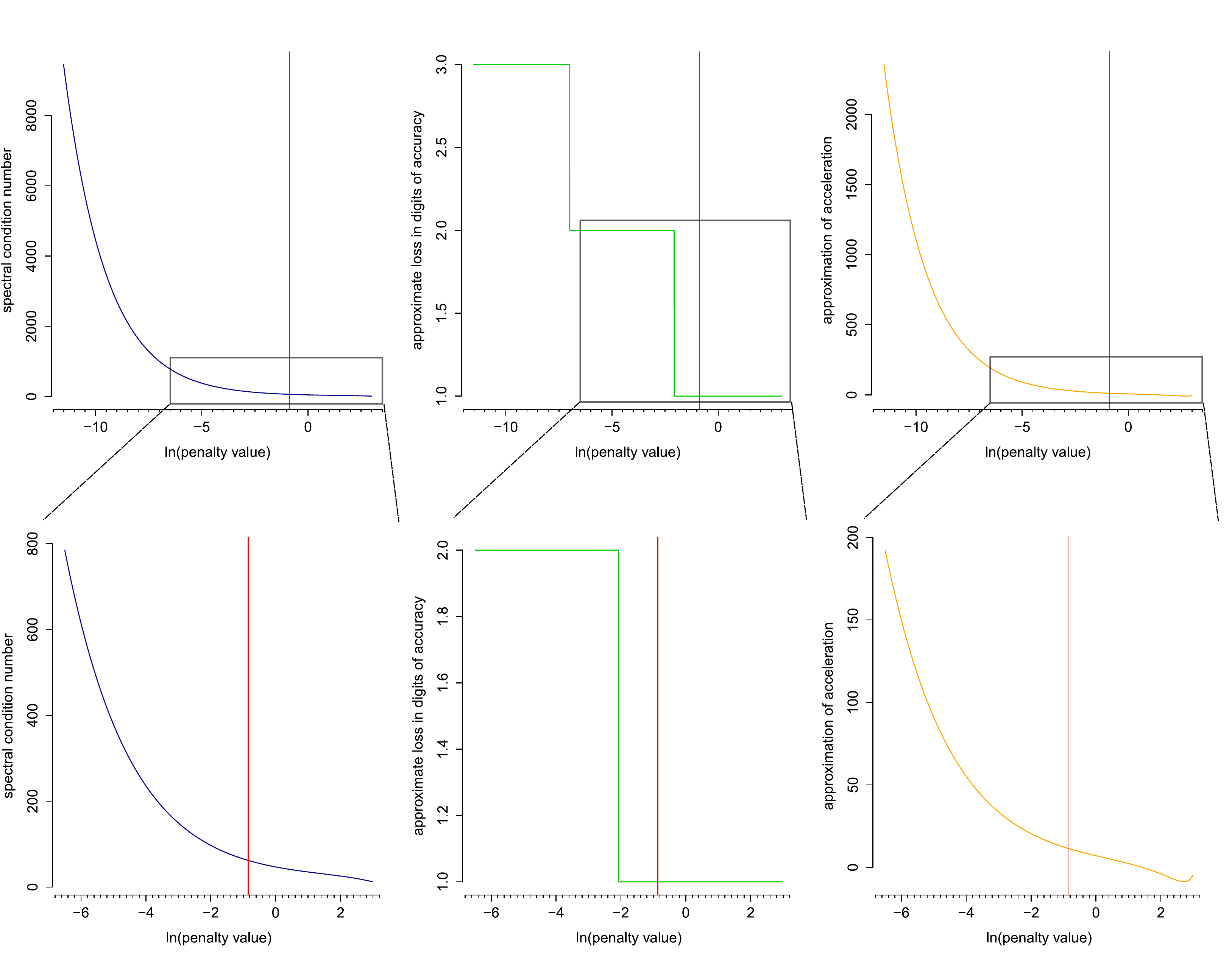}
  \caption{
The left-hand panels give the basic spectral condition number plot.
The middle and right-hand panels exemplify the interpretational aids to the basic plot:
the approximate loss in digits of accuracy (middle panel) and the approximation of the acceleration along the curve in the basic plot (right-hand panel).
The top panels give the basic condition number plot and its interpretational aids for the domain $\lambda_{a} \in [\num{1e-5},20]$.
The bottom panels zoom in on the boxed areas.
The boxed areas cover a domain of well-conditionedness according to the heuristically chosen minimal penalty-value: $\lambda_{a} \in [\exp(-6.5),20]$.
The red vertical line indicates the value of the penalty that was chosen as optimal by the root-finding LOOCV procedure ($.422$).
  }
  \label{Illus2WAIDS}
\end{figure}

The estimator of choice is again (3) of the main text where we replace $\hat{\mathbf{\Sigma}}$ with the sample correlation matrix $\mathbf{R}$.
The target matrix is chosen as $\mathbf{T} = \mathbf{I}_{p}$, such that a regularized correlation matrix ensues.
Again, the aLOOCV procedure was tried first in finding an optimal value for $\lambda_a$ under the given target and data settings.
We searched for the optimal value $\lambda_{a}^{*}$ in the domain $\lambda_{a} \in [\num{1e-5},20]$ with $10,000$ log-equidistant steps along this domain.
Again, the procedure pointed to $\num{1e-5}$ as being the optimal value for the penalty (in the chosen domain), which seems low given the $p/n$ ratio of the data.
The condition number plot (covering the same penalty-domain considered by the aLOOCV procedure) indeed indicates that the precision estimate at $\lambda_{a} = \num{1e-5}$ is not well-conditioned in the sense of the Heuristic Definition, exhibiting a condition number of approximately $9,456$ (Figure \ref{Illus2WAIDS}).
A reasonable minimal penalty-value (in accordance with the Heuristic Definition) can be found at approximately $\exp(-6.5)$, at which $\mathcal{C}_{2}[\hat{\mathbf{\Sigma}}^{a}(\exp(-6.5))] \approx 785.01$.
This reasonable minimal value is subsequently used to constrain the search-domain to the region of well-conditionedness.
A root-finding (by the Brent algorithm \citep{Brent}) LOOCV procedure is then told to search for the optimal value $\lambda_{a}^{*}$ in the domain $\lambda_{a} \in [\exp(-6.5),20]$.
The optimal penalty-value is found at $.422$.
At this value, indicated by the red vertical line in Figure \ref{Illus2WAIDS}, $\mathcal{C}_{2}[\hat{\mathbf{\Sigma}}^{a}(.422)] \approx 62.39$.

The Kaiser-Meyer-Olkin index \citep{2ndJIFF} of .98 indicates that a reasonable proportion of variance among the variables might be common variance and, hence, that $\hat{\mathbf{\Sigma}}^{a}(.422)$ is suitable for a factor analysis. The regularized estimate $\hat{\mathbf{\Sigma}}^{a}(.422)$ is used in further analysis.

\subsection*{2.4. Further analysis}
\begin{figure}
  \begin{minipage}[b]{0.60\linewidth}
    \centering
    \includegraphics[scale = .45]{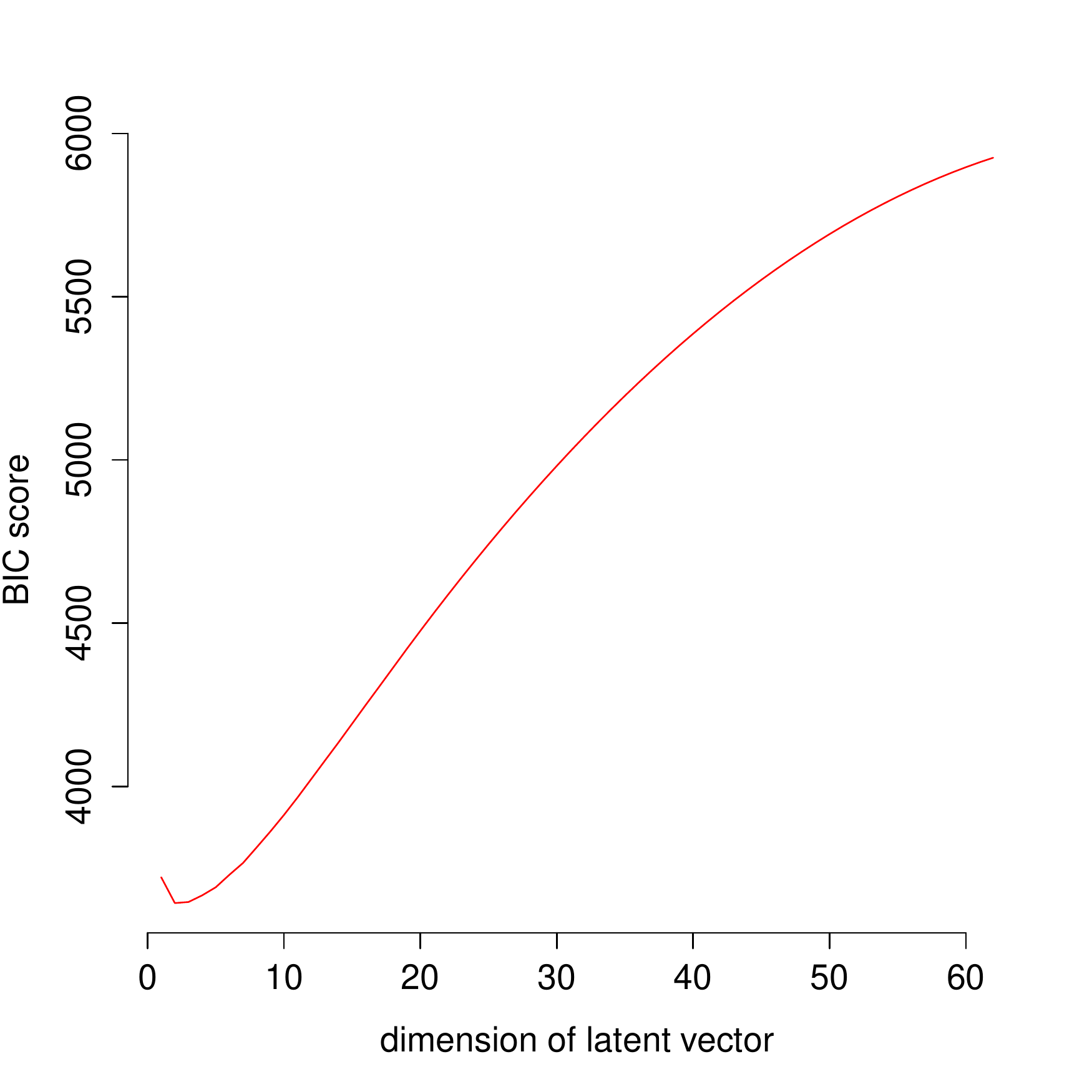}
    \par\vspace{0pt}
  \end{minipage}%
  \begin{minipage}[b]{0.36\linewidth}
    \centering%
\begin{tabular}{cc}
  \hline\hline \\
  $m$ & BIC \\ \hline \\
  1 & $3,721.762$ \\
  2 & $3,642.678$ \\
  3 & $3,645.902$ \\
  4 & $3,666.222$ \\
  5 & $3,691.069$ \\
  \hline
\\
\\
\\
\\
\\
\end{tabular}
    \par\vspace{0pt}
  \end{minipage}
\caption{BIC scores for various dimensions of the latent vector.
The left-hand panel gives the trace of the BIC scores for all dimensions of the latent vector
allowed by the bound $p(p+1)/2 - \eta \geq 0$. The right-hand panel gives a table with
BIC scores for $m = 1, \ldots, 5$. The $m = 2$ solution is to be preferred according to the BIC.}
\label{BICsolution}
\end{figure}

The dimension of the factor solution is unknown.
Hence, the optimal factor dimension needs to be determined in conjunction with the estimation of the model.
Now, let $\hat{\mathbf{\Sigma}}_{m} = \hat{\mathbf{\Lambda}}_{m}\hat{\mathbf{\Lambda}}_{m}^{\mathrm{T}} + \hat{\mathbf{\Psi}}$ denote the MLE solution under $m$ factors.
Then, we determine the optimal dimension (contingent upon the MLE solutions) using the Bayesian Information Criterion (BIC; \citep{BIC}), which, for the problem at hand, amounts to:
\begin{equation}\nonumber
n \left\{ p \ln(2\pi) + \ln|\hat{\mathbf{\Sigma}}_{m}| + \mbox{tr}\left( \hat{\mathbf{\Sigma}}_{m}^{-1}\hat{\mathbf{\Sigma}}^{a}(.422) \right) \right\} + \ln(n)\eta,
\end{equation}
where $\eta = p(m + 1) - m(m - 1)/2$ indicates the free parameters in the model.
Figure \ref{BICsolution} gives the BIC scores for each dimension allowed by the existence condition $p(p+1)/2 - \eta \geq 0$.
The solution with the lowest BIC score is deemed optimal, indicating that $m = 2$ is the preferred solution.

The specified FA model is inherently underidentified.
Assume $\mathbf{H} \in \mathbb{R}^{m \times m}$ is an arbitrary orthogonal matrix.
Considering the implied covariance structure of the observed data we may write:
\begin{equation}\nonumber
\mathbf{\Gamma}\mathbf{\Gamma}^{\mathrm{T}} + \mathbf{\Psi} = (\mathbf{\Gamma}\mathbf{H})(\mathbf{\Gamma}\mathbf{H})^{\mathrm{T}} + \mathbf{\Psi}.
\end{equation}
This equality implies that, given $\mathbf{\Psi}$, there is an infinite number of alternative loading matrices
that generate the same covariance structure as $\mathbf{\Gamma}$.
Thus, in any solution, $\mathbf{\Gamma}$ can be made to satisfy $m(m - 1)/2$ additional conditions, and, hence, the structure of the existence condition given above.
Naturally, any estimation method then requires a minimum of $m(m - 1)/2$ restrictions on $\mathbf{\Gamma}$ to attain uniqueness (up to possibly polarity reversals in the columns of $\mathbf{\Gamma}$).
In MLE this is achieved by requiring that $\mathbf{\Gamma}^{\mathrm{T}}\mathbf{\Psi}^{-1}\mathbf{\Gamma}$ be diagonal along with an order-condition on its diagonal elements.
As this is a convenience solution that has no direct interpretational meaning, usually a post-hoc rotation is applied, whence estimation is settled, to enhance interpretation.
Here, the Varimax \cite{Varimax} rotation to an orthogonal simple structure was ultimately used as oblique rotation (that allows for factor correlations) indicated a near-zero correlation between the two retained latent factors (note that any orthogonal representation has equivalent oblique representations).

\begin{figure}[tbp]
  \centering
  \includegraphics[width=\textwidth]{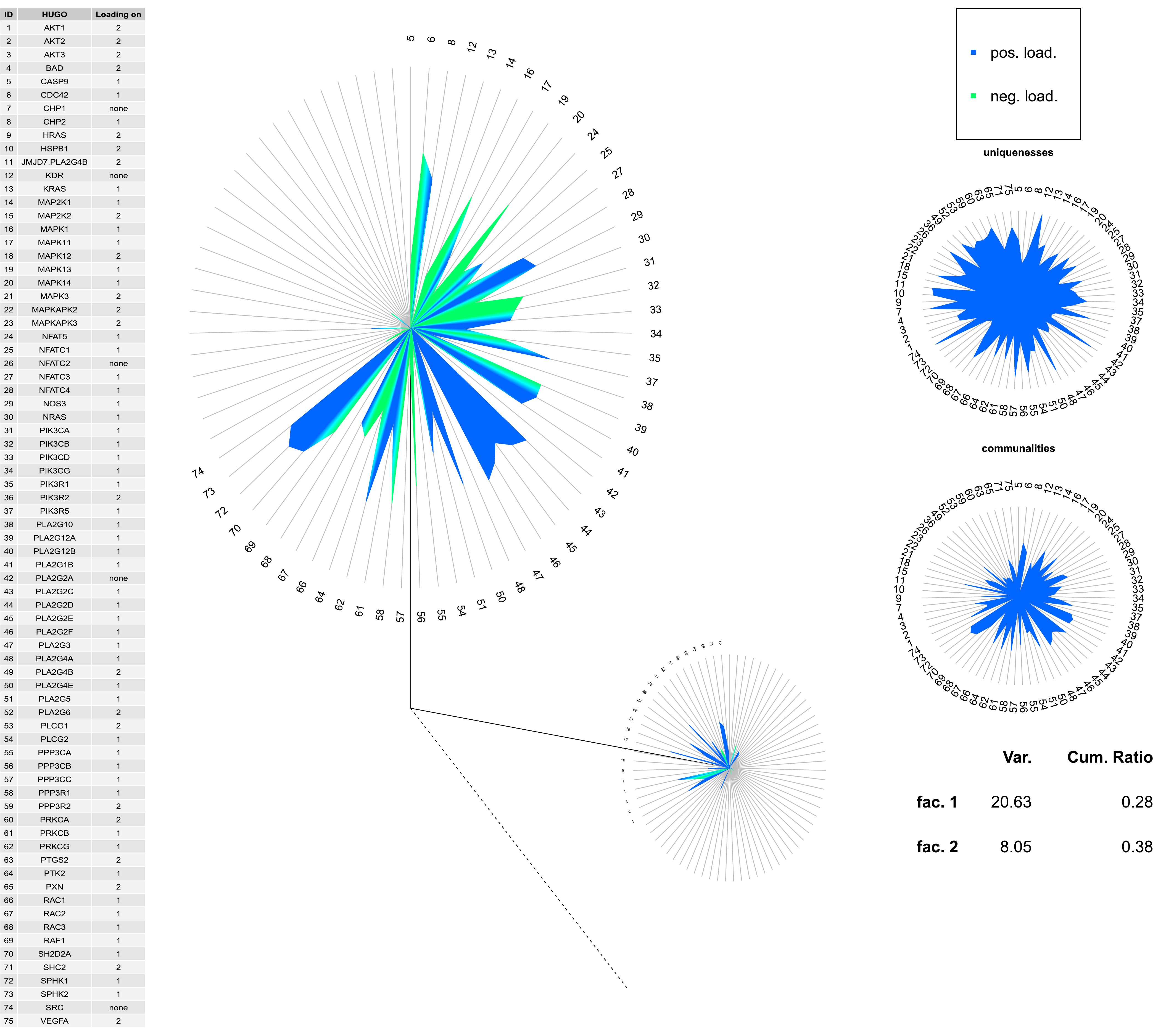}
  \caption{
    Dandelion plot of the factor solution.
    Each central solid line represents a factor and is connected to a star plot (also known as a Kiviat diagram).
    Each spoke in each of the star plots then represents a variable.
    The length of the spokes corresponds to the maximum magnitude of each loading (which is unity).
    The extension of the polygon along each spoke then indicates the magnitude of a factor loading in the obtained solution, with the sign of the corresponding loading represented by color coding:
    green for a negative loading and blue for a positive loading.
    The representation suppresses absolute loadings lower than .3 for enhanced visualization.
    The angle between the solid lines corresponds to the amount of variance explained by the first factor (approximately 28\%).
    The dashed line represents the cumulative percentage of variance explained by all retained factors (approximately 38\% in this case).
    The star plots at the far-right then represent the magnitudes of the communalities $\kappa_j = \sum_j \gamma_{jk}^{2}$ and the uniquenesses $\psi_j = 1 - \kappa_j$.
    Variables are represented by index numbers.
    The table at the far-left gives the corresponding HUGO-curated gene names.
    See \cite{Dandelion} for more information on the Dandelion plot.
  }
  \label{Dandelion}
\end{figure}

The final factor solution is represented in  Figure \ref{Dandelion}.
This Dandelion plot \cite{Dandelion} visualizes the magnitudes of the elements of $\hat{\mathbf{\Lambda}}$ and $\hat{\mathbf{\Psi}}$ and their relation to the two retained factors.
The latent factors are taken to represent biochemical {\nobreak subpwathways}.
Table \ref{FAgenes} characterizes the factors according to their constituting genes that achieve the highest absolute loading.
These genes are furthermore associated with VEGF-signaling subpathways as indicated by the WikiPathways database \citep{WikiPathways}.
Factor 1 consists mostly of MAPK-related genes.
In addition, genes related to the glycerol phospholipid biosynthesis pathway load on factor 1.
The MAPK pathway has been associated with lipid homeostase in yeast \cite{MAPKlips} and ``phospholipid composition and synthesis are similar in yeast and mammalian cells" \citep{Phospholips}.
Genes with pathway association indicated by `-' in Table \ref{FAgenes} (PIK3R5 and SPHK2) could not be associated to VEGF-subpathways according to the WikiPathways database.
PIK3R5 is, however, related to the MAPK-signaling pathway according to the Ingenuity Target Explorer \citep{QIAGEN}.
Factor 2 consists mostly of genes related to the PI3K/AKT-signaling and prostate cancer pathways.
Moreover, this factor also includes general VEGF-signaling pathway genes that are conducive in activating the MAPK and PI3K/AKT cascade.
These results suggest that the compositional subpathway structure of the VEGF-signaling pathway in metastatic prostate cancer can still be characterized
as consisting of the MAPK and PI3K-AKT cascades.
Hence, VEGF-pathway degerulation in metastatic prostate cancer is more likely characterized by intricate changes in its known subpathways than by the loss of these (normal) subpathways or the gain of (abnormal) subpathways.

\begin{table}[]
\caption{Top genes per factor according to absolute factor loadings.
`HUGO' refers to Human Genome Organization Gene Nomenclature Committee (HGNC) curated gene names.
`Pathway' refers to the VEGF-subpathway a certain gene belongs to according to the WikiPathways database \citep{WikiPathways}.}
\centering
\label{FAgenes}
\begin{tabular}{crl}
 \hline\hline \\
 Factor & HUGO Symbol & Pathway \\
 \hline \\
      \textbf{1}   & CDC42   & MAPK-signaling  \\
         & MAPK14  & MAPK-signaling  \\
         & PIK3R5  & -  \\
         & PLA2G2C & Glycerol phospholipid biosynthesis  \\
         & PLA2G3  & Glycerol phospholipid biosynthesis  \\
         & PLA2G4E & Glycerol phospholipid biosynthesis  \\
         & PPP3R1  & MAPK-signaling  \\
         & PRKCG   & MAPK-signaling  \\
         & SPHK1   & VEGF-signaling  \\
         & SPHK2   & -  \\
         &         &    \\
      \textbf{2}   & AKT1    & PI3K/AKT-signaling in cancer  \\
         & AKT2    & AKT-signaling / VEGF-signaling  \\
         & AKT3    & AKT-signaling / VEGF-signaling  \\
         & BAD     & AKT-signaling / Prostate cancer \\
         & MAP2K2  & MAPK-signaling / Prostate cancer \\
         & MAPKAPK2& MAPK-signaling / Prostate cancer \\
         & PIK3R2  & AKT-signaling / VEGF-signaling \\
         & PLA2G6  & Glycerol phospholipid biosynthesis \\
         & PXN     & VEGF-signaling / Prostate cancer \\
         & SHC2    & VEGF-signaling \\
 \hline
\end{tabular}
\end{table}

\newpage
\section*{3. Benchmark}
A benchmarking exercise was conducted to get an indication of the execution time of the \texttt{CNplot} function.
The sample size $n$ is not a factor in the execution time as the function operates on the covariance (or precision) matrix.
Hence, time complexity hinges on the dimension $p$ and the number of steps $S$ taken along the (specified) domain of the penalty-parameter.
Execution time was thus evaluated for various combinations of $p$ and $S$.

\begin{sloppypar}
Timing evaluations were done for all ridge estimators mentioned in Section 2.1 of the main text and all combinations of the elements in $p \in \{125,250,500,1000\}$ and $S \in \{125,250,500,1000\}$.
The basic condition number plot was produced 50 times (on simulated covariance matrices) for each combination of estimator, $p$, and $S$.
In addition, both a rotation equivariant situation (using a scalar target) and a rotation non-equivariant situation (using a non-scalar target) were considered for estimators (1) and (3) of the main text (note that estimator (2) is always rotation equivariant).
All execution times of the \texttt{CNplot} call in \textsf{R} were evaluated with the \texttt{microbenchmark} package \citep{BenchMark}.
All timings were carried out on a Intel$^{\circledR}$ Core$^{\mathrm{TM}}$ i7-4702MQ 2.2 GHz processor.
The results, stated as median runtimes in seconds, are given in Tables \ref{BMarkREalt} to \ref{BMarkREarchII}.
The code used in this benchmarking exercise can be found in Listing \ref{LIST:Bench} in Section 4 of this Supplement.
\end{sloppypar}

Tables \ref{BMarkREalt} to \ref{BMarkREarchII} indicate that, in general, the condition number plot can be produced quite swiftly.
As stated in the main text: in rotation equivariant situations only a single spectral decomposition is required to obtain the complete solution path of the condition number.
Hence, the time complexity of the \texttt{CNplot} call in these situations is approximately $\mathcal{O}(p^{3})$ (worst-case time complexity of a spectral decomposition) as, after the required spectral decomposition, the solution path can be obtained in linear time only. Increasing the number of steps $S$ along the penalty-domain thus comes at little additional computational cost.
For the rotation non-equivariant setting the time complexity is approximately $\mathcal{O}(Sp^{3})$ as a spectral decomposition is required for all $s = 1, \ldots, S$.
The runtime of the condition number plot function under Estimator (3) of the main text is then somewhat longer than the corresponding runtime under Estimator (1).
This is because the spectral decomposition in the former situation is also used for computing the (relatively expensive) matrix square root.
As $S$ acts as a scaling factor in rotation non-equivariant settings the computation time of the plot can be reduced by coarsening the search-grid along the penalty-domain.

\begin{table}[]
\caption{Benchmarking results for various values of $p$ and $S$ for estimator (3) of the main text.
The results respresent median runtimes in seconds.}
\centering
\label{BMarkREalt}
\begin{tabular}{l|rrrr}
 \hline \hline \\
 & $p = 125$ & $p = 250$ & $p = 500$ & $p = 1000$ \\
 \hline
   \rowcolor{gray!20} & \multicolumn{4}{c}{Rotation equivariant setting} \\
   $S = 125$     & .023  & .044 & .222 & .998 \\
   $S = 250$     & .024  & .046 & .218 & .965 \\
   $S = 500$     & .027  & .050 & .221 & 1.122 \\
   $S = 1000$    & .035  & .061 & .233 & 1.188 \\
\rowcolor{gray!20} & \multicolumn{4}{c}{Rotation non-equivariant setting}\\
   $S = 125$     &  .689  &  3.267 &  18.801 & 139.982 \\
   $S = 250$     & 1.315  &  6.661 &  39.022 & 278.454 \\
   $S = 500$     & 2.703  & 13.418 &  76.489 & 553.812 \\
   $S = 1000$    & 4.995  & 26.112 & 152.427 & 1,106.002 \\
 \hline
\end{tabular}
\end{table}

\begin{table}[]
\caption{Benchmarking results for various values of $p$ and $S$ for estimator (1) of the main text.
The results respresent median runtimes in seconds.}
\centering
\label{BMarkREarchI}
\begin{tabular}{l|rrrr}
 \hline \hline \\
 & $p = 125$ & $p = 250$ & $p = 500$ & $p = 1000$ \\
  \hline
   \rowcolor{gray!20} & \multicolumn{4}{c}{Rotation equivariant setting} \\
   $S = 125$     & .023  & .044 & .214 & 1.179 \\
   $S = 250$     & .024  & .048 & .218 & 1.227 \\
   $S = 500$     & .027  & .054 & .219 & 1.032 \\
   $S = 1000$    & .031  & .054 & .222 & 1.046 \\
\rowcolor{gray!20} & \multicolumn{4}{c}{Rotation non-equivariant setting}\\
   $S = 125$     & .339   & 1.448  & 9.701  & 65.988 \\
   $S = 250$     & .590   & 2.849  & 18.128 & 129.262 \\
   $S = 500$     & 1.090  & 5.385  & 35.188 & 281.583 \\
   $S = 1000$    & 2.366  & 12.199 & 75.222 & 522.910 \\
 \hline
\end{tabular}
\end{table}

\begin{table}[]
\caption{Benchmarking results for various values of $p$ and $S$ for estimator (2) of the main text.
The results respresent median runtimes in seconds.}
\centering
\label{BMarkREarchII}
\begin{tabular}{l|rrrr}
 \hline \hline \\
 & $p = 125$ & $p = 250$ & $p = 500$ & $p = 1000$ \\
 \hline
   \rowcolor{gray!20} & \multicolumn{4}{c}{Rotation equivariant setting} \\
   $S = 125$     & .026  & .056 & .281 & 1.606 \\
   $S = 250$     & .026  & .059 & .282 & 1.597 \\
   $S = 500$     & .027  & .058 & .261 & 1.600 \\
   $S = 1000$    & .030  & .060 & .280 & 1.689 \\
 \hline
\end{tabular}
\end{table}

To compare the runtimes for the \texttt{CNplot} call we also conducted timing evaluations for the approximate leave-one-out cross-validation (aLOOCV) and root-finding LOOCV procedures as used in the main text (implemented in \texttt{rags2ridges}).
Time complexity for the root-finding LOOCV procedure hinges on the dimension $p$ and the sample size $n$.
In contrast, time complexity for for the aLOOCV procedure is dependent on $p$, $n$, and $S$.
Timings were done for all combinations of the elements in $p \in \{125,250\}$, $n \in \{100,200\}$ and (when relevant) $S \in \{125,250\}$.
The basic condition number plot was produced 50 times (on simulated data of dimension $n \times p$) for each combination of $p$, $n$ and (possibly) $S$.
For the root-finding LOOCV procedure both a rotation equivariant situation and a rotation non-equivariant situation were considered.
Ridge estimator (3) of the main text was used for all timing evaluations.
The results, stated as median runtimes in seconds, are given in Tables \ref{BMarkRFloocv} and \ref{BMarkaLOOCV}.
The code used in this exercise can also be found in Listing \ref{LIST:Bench} in Section 4 of this Supplement.

Tables \ref{BMarkRFloocv} and \ref{BMarkaLOOCV} indicate that, even under these relatively light settings, the runtimes for the aLOOCV and root-finding LOOCV procedures far exceed the runtime of (corresponding) calls to \texttt{CNplot}.
The runtime for the root-finding LOOCV procedure is (given $p$) exacerbated for larger sample sizes.
The runtime for the aLOOCV procedure is (given $p$) exacerbated for larger samples sizes, finer search-grids, and (not shown) situations of rotation non-equivariance.

\begin{table}[]
\caption{Benchmarking results for various values of $p$ and $n$ for the root-finding LOOCV procedure.
The ridge estimator considered is given in equation (3) of the main text.
The results respresent median runtimes in seconds.}
\centering
\label{BMarkRFloocv}
\begin{tabular}{l|rr}
 \hline \hline \\
 & $p = 125$ & $p = 250$ \\
 \hline
   \rowcolor{gray!20} & \multicolumn{2}{c}{Rotation equivariant setting} \\
   $n = 100$     & 36.843  & 186.662 \\
   $n = 200$     & 83.932  & 438.080 \\
\rowcolor{gray!20} & \multicolumn{2}{c}{Rotation non-equivariant setting}\\
   $n = 100$     & 37.151  & 186.780 \\
   $n = 200$     & 84.335  & 439.607 \\
 \hline
\end{tabular}
\end{table}

\begin{table}[]
\caption{Benchmarking results for various values of $p$, $n$, and $S$ for the approximate LOOCV procedure.
The ridge estimator considered is given in equation (3) of the main text.
The results respresent median runtimes in seconds.}
\centering
\label{BMarkaLOOCV}
\begin{tabular}{l|rr}
 \hline \hline \\
 & $p = 125$ & $p = 250$ \\
 \hline
   \rowcolor{gray!20} & \multicolumn{2}{c}{$n = 100$, Rotation equivariant setting} \\
   $S = 125$     &  53.059  & 393.309 \\
   $S = 250$     & 100.733  & 802.741\\
   \rowcolor{gray!20} & \multicolumn{2}{c}{$n = 200$, Rotation equivariant setting} \\
   $S = 125$     & 105.372  &  779.881 \\
   $S = 250$     & 212.129  & 1561.064\\
 \hline
\end{tabular}
\end{table}

\newpage
\section*{4. \textsf{R} Codes}

The \textsf{R} libraries needed to run the code snippets below can be found in Listing \ref{LIST:Packages}:

\begin{lstlisting}[language = R, caption = Packages and dependencies, label = LIST:Packages]
####################################################
##--------------------------------------------------
## Packages and dependencies
##--------------------------------------------------
####################################################

## R libraries
library(biomaRt)
library(cgdsr)
library(KEGG.db)
library(microbenchmark)
library(psych)
library(DandEFA)
library(gridExtra)
library(plyr)
library(rags2ridges)
\end{lstlisting}

\bigskip
The code in Listing \ref{LIST:Fig1} will produce Figure 1 of the main text:

\begin{lstlisting}[language = R, caption = Code for Figure 1, label = LIST:Fig1]
####################################################
##--------------------------------------------------
## First example of copy number Plot
##--------------------------------------------------
####################################################

## Obtain covariance matrix on p > n data
p = 100
n = 25
set.seed(333)
X = matrix(rnorm(n*p), nrow = n, ncol = p)
Cx <- covML(X)

## Obtain basic spectral condition number plot
CNplot(Cx, lambdaMin = .001, lambdaMax = 10, step = 1000,
       type = "ArchII")

## Condition number at exp(-3)
EVs <- eigen(ridgeP(Cx, lambda = exp(-3),
                    type = "ArchII"))$values
Cn  <- EVs[1]/EVs[ncol(Cx)]; Cn
\end{lstlisting}

\bigskip
The code in Listing \ref{LIST:Fig2} will produce (the components of) Figure 2 of the main text:

\begin{lstlisting}[language = R, caption = Code for Figure 2, label = LIST:Fig2]
####################################################
##--------------------------------------------------
## Example copy number Plot with interpretational aids
##--------------------------------------------------
####################################################

## Spectral condition number plot with interpretational aids
CNplot(Cx, lambdaMin = .00001, lambdaMax = 1000, step = 2000,
       Iaids = TRUE, type = "Alt",
       target = default.target(Cx, type = "DEPV"))

## Zoom
CNplot(Cx, lambdaMin = .01, lambdaMax = 1000, step = 2000,
       Iaids = TRUE, type = "Alt",
       target = default.target(Cx, type = "DEPV"))
\end{lstlisting}

\bigskip
The code in Listing \ref{LIST:EX1} was used for Section 4 of the main text:

\begin{lstlisting}[language = R, caption = Code for Illustration 1, label = LIST:EX1]
####################################################
##--------------------------------------------------
## Illustration 1: Kidney cancer data
##--------------------------------------------------
####################################################

####################################################
## Probe MSKCC Cancer Genomics Data Server for data
####################################################

## Get list all human genes
ensembl = useMart("ENSEMBL_MART_ENSEMBL",
                  dataset = "hsapiens_gene_ensembl",
                  host="www.ensembl.org")
geneList <- getBM(attributes = c("hgnc_symbol", "entrezgene"),
                  mart = ensembl)
geneList <- geneList[!is.na(geneList[,2]),]

## Obtain entrez IDs of genes that map to Hedgehog signaling pathway
kegg2entrez <- as.list(KEGGPATHID2EXTID)
entrezIDs   <- as.numeric(kegg2entrez[which(names(kegg2entrez) 
                          %in% "hsa04340")][[1]])
entrez2name <- match(entrezIDs, geneList[,2])
geneList    <- geneList[entrez2name[!is.na(entrez2name)],]

## Specify data set details
mskccDB     <- CGDS("http://www.cbioportal.org/public-portal/")
studies     <- getCancerStudies(mskccDB)
cancerStudy <- "kirc_tcga_pub"
caseList    <- getCaseLists(mskccDB, cancerStudy)
caseList    <- getCaseLists(mskccDB, cancerStudy)[3,1]
mygeneticprofile = getGeneticProfiles(mskccDB,cancerStudy)
mrnaProf    <- "kirc_tcga_pub_rna_seq_v2_mrna_median_Zscores"

## Extract data
Y <- getProfileData(mskccDB, geneList[,1], mrnaProf, caseList)
Y <- as.matrix(Y)

## Filter no-data samples and genes and scale data
sRemove <- which(rowSums(is.na(Y)) > ncol(Y)/2); sRemove
gRemove <- which(colSums(is.na(Y)) > 0); gRemove
Y       <- scale(Y, center = TRUE, scale = TRUE)


####################################################
## Regularize the precision matrix
####################################################

## Approximate LOOCV
## Chooses very small penalty
aLOOCVres <- optPenalty.aLOOCV(Y, 1e-05, 20, 10000, 
                               type = "Alt", cor = TRUE, 
                               target = default.target(cor(Y), 
                                                       type = "DUPV"))

## Condition number plot
## aLOOCV penalty indeed too small
## Can give idea good heuristic value for penalty
CNplot(cor(Y), 1e-05, 20, 5000, type = "Alt",
       target = default.target(cor(Y), type = "DUPV"))

## Perform LOOCV (with Brent) and restrict search space on basis CnPlot
LOOCVres  <- optPenalty.LOOCVauto(Y, exp(-6), 20, 
                                  type = "Alt", cor = TRUE,
                                  target = default.target(cor(Y), 
                                                       type = "DUPV"))
LOOCVres$optLambda

## Condition number plot with optimal LOOCV-determined penalty indicated
## 'Optimal' approx. LOOCV-determined penalty is also indicated
CNplot(cor(Y), 1e-05, 20, 5000, type = "Alt",
       target = default.target(cor(Y), type = "DUPV"),
       vertical = TRUE, value = LOOCVres$optLambda)
abline(v = log(aLOOCVres$optLambda), col = "green")


####################################################
## Assessment condition numbers
####################################################

## Condition number at optimal value indicated by aLOOCV
EVs <- eigen(ridgeP(cor(Y), lambda = 1e-05, type = "Alt",
                    target = default.target(cor(Y), 
                                            type = "DUPV")))$values
Cn  <- EVs[1]/EVs[ncol(Y)]; Cn

## Condition number at heuristic value
EVs <- eigen(ridgeP(cor(Y), lambda = exp(-6), type = "Alt",
                    target = default.target(cor(Y), 
                                            type = "DUPV")))$values
Cn  <- EVs[1]/EVs[ncol(Y)]; Cn

## Condition number at optimal value indicated by LOOCV
EVs <- eigen(ridgeP(cor(Y), lambda = LOOCVres$optLambda, type = "Alt",
                    target = default.target(cor(Y), 
                                            type = "DUPV")))$values
Cn  <- EVs[1]/EVs[ncol(Y)]; Cn


####################################################
## Downstream graphical modeling
####################################################

Pp0 <- sparsify(LOOCVres$optPrec, "localFDR", FDRcut = .8)
edgeHeat(Pp0$sparseParCor)
Ugraph(Pp0$sparseParCor, type = "fancy",
       lay = "layout_with_fr",
       Vcolor = "white", VBcolor = "black",
       Vcex = .5, cut = .Machine$double.xmin,
       prune = T)
\end{lstlisting}

\bigskip
The code in Listing \ref{LIST:EX2} was used for the second illustration contained in Section 2 of this supplement:

\begin{lstlisting}[language = R, caption = Code for Illustration 2, label = LIST:EX2]
####################################################
##--------------------------------------------------
## Illustration 2: Prostate cancer data
##--------------------------------------------------
####################################################

####################################################
## Probe MSKCC Cancer Genomics Data Server for data
####################################################

## Get list all human genes
ensembl = useMart("ENSEMBL_MART_ENSEMBL",
                  dataset = "hsapiens_gene_ensembl",
                  host="www.ensembl.org")
geneList <- getBM(attributes = c("hgnc_symbol", "entrezgene"),
                  mart = ensembl)
geneList <- geneList[!is.na(geneList[,2]),]

## Obtain entrez IDs of genes that map to VEGF signaling pathway
kegg2entrez <- as.list(KEGGPATHID2EXTID)
entrezIDs   <- as.numeric(kegg2entrez[which(names(kegg2entrez)
                                            %in% "hsa04370")][[1]])
entrez2name <- match(entrezIDs, geneList[,2])
geneList    <- geneList[entrez2name[!is.na(entrez2name)],]

## Specify data set details
mskccDB     <- CGDS("http://www.cbioportal.org/public-portal/")
studies     <- getCancerStudies(mskccDB)
cancerStudy <- "prad_mskcc"
caseList    <- getCaseLists(mskccDB, cancerStudy)[15,1]
mygeneticprofile = getGeneticProfiles(mskccDB,cancerStudy)
mrnaProf    <- "prad_mskcc_mrna"

## Extract data
Y2 <- getProfileData(mskccDB, geneList[,1], mrnaProf, caseList)
Y2 <- as.matrix(Y2)

## Filter no-data samples and genes and scale data
sRemove <- which(rowSums(is.na(Y2)) > ncol(Y2)/2); sRemove
gRemove <- which(colSums(is.na(Y2)) > 0); gRemove
Y2      <- scale(Y2, center = TRUE, scale = TRUE)


####################################################
## Regularize the precision matrix
####################################################

## Approximate LOOCV
## Chooses very small penalty
aLOOCVres <- optPenalty.aLOOCV(Y2, 1e-05, 20, 10000, 
                               type = "Alt", cor = TRUE,
                               target = default.target(cor(Y2),
                                                       type = "DUPV"))
aLOOCVres$optLambda

## Condition number plot
## aLOOCV penalty indeed too small
## Can give idea good heuristic value for penalty
CNplot(cor(Y2), 1e-05, 20, 5000, type = "Alt",
       target = default.target(cor(Y2), type = "DUPV"))

## Perform LOOCV (with Brent) and restrict search space on basis CnPlot
LOOCVres  <- optPenalty.LOOCVauto(Y2, exp(-6.5), 20, 
                                  type = "Alt", cor = TRUE,
                                  target = default.target(cor(Y2),
                                                       type = "DUPV"))
LOOCVres$optLambda

## Condition number plot with optimal LOOCV-determined penalty indicated
CNplot(cor(Y2), 1e-05, 20, 5000, type = "Alt",
       target = default.target(cor(Y2), type = "DUPV"),
       vertical = TRUE, Iaids = TRUE, value = LOOCVres$optLambda)
CNplot(cor(Y2), exp(-6.5), 20, 5000, type = "Alt",
       target = default.target(cor(Y2), type = "DUPV"),
       vertical = TRUE, Iaids = TRUE, value = LOOCVres$optLambda)


####################################################
## Assessment condition numbers
####################################################

## Condition number at optimal value indicated by aLOOCV
EVs <- eigen(ridgeP(cor(Y2), lambda = aLOOCVres$optLambda,
                    type = "Alt",
                    target = default.target(cor(Y2),
                                            type = "DUPV")))$values
Cn  <- EVs[1]/EVs[ncol(Y2)]; Cn

## Condition number at heuristic value
EVs <- eigen(ridgeP(cor(Y2), lambda = exp(-6.5), type = "Alt",
                    target = default.target(cor(Y2),
                                            type = "DUPV")))$values
Cn  <- EVs[1]/EVs[ncol(Y2)]; Cn

## Condition number at optimal value indicated by LOOCV
EVs <- eigen(ridgeP(cor(Y2), lambda = LOOCVres$optLambda,
                    type = "Alt",
                    target = default.target(cor(Y2),
                                            type = "DUPV")))$values
Cn  <- EVs[1]/EVs[ncol(Y2)]; Cn


####################################################
## Downstream factor analytic modeling
####################################################

##--------------------------------------
## Obtain regularized correlation matrix
## Assess factorability
##--------------------------------------
R = cov2cor(solve(LOOCVres$optPrec))
KMO(R)

##--------------------------------------
## Determine dimension latent vector
##--------------------------------------

## Function
BICfacM <- function(S, n, m){
  ######################################
  ## S > (regularized) covariance or
  ##      correlation matrix
  ## n > sample size
  ## m > desired number of factors
  ######################################

  ## Preliminaries
  p    <- ncol(S)
  fit  <- factanal(factors = m, covmat = S, rotation = "none")
  loadings <- fit$loadings[1:p,]
  Uniqueness <- diag(fit$uniquenesses)
  Sfit <- loadings %*% t(loadings) + Uniqueness

  ## Calculate BIC
  fit <- n * (p*log(2*pi) + log(det(Sfit)) +
         sum(diag(solve(Sfit) %*% S)))
  penalty <- p*(m+1) - (m*(m-1))/2
  BIC <- fit + penalty

  ## Return
  return(BIC)
}

## Ledermann bound
p    <- ncol(R)
mmax <- floor((2*p+1 - sqrt(8*p+1))/2)

## Determine optimal dimension
BIC <- numeric()
for(m in 1:(mmax - 1)){
  BIC[m] <- BICfacM(R, n = 19, m = m)
}; BIC

## Plot
dims <- seq(1, (mmax - 1), 1)
plot(dims, BIC, axes = FALSE, type = "l",
     col = "red", xlab = "dimension of latent vector",
     ylab = "BIC score")
axis(2, ylim = c(min(BIC),max(BIC)), col = "black", lwd = 1)
axis(1, xlim = c(0,(mmax - 1)), col = "black", lwd = 1, tick = TRUE)

##--------------------------------------
## Fit under optimal dimension
##--------------------------------------
fit <- factanal(factors = 2, covmat = R, rotation = "promax")
print(fit, digits = 2, cutoff = .3, sort = FALSE)
fit <- factanal(factors = 2, covmat = R, rotation = "varimax")
print(fit, digits = 2, cutoff = .3, sort = TRUE)

##--------------------------------------
## Visualizing solution
##--------------------------------------

## Dandelion plot
Loading <- character()
for (i in 1:nrow(fit$loadings)){
  if (abs(fit$loadings[i,1]) >= .3 & abs(fit$loadings[i,1]) > abs(fit$loadings[i,2])){
    Loading[i] <- "1"
  }
  if (abs(fit$loadings[i,2]) >= .3 & abs(fit$loadings[i,2]) > abs(fit$loadings[i,1])){
    Loading[i] <- "2"
  }
  if (abs(fit$loadings[i,1]) == abs(fit$loadings[i,2])){
    Loading[i] <- "both"
  }
  if (abs(fit$loadings[i,1]) < .3 & abs(fit$loadings[i,2]) < .3){
    Loading[i] <- "none"
  }
}
Names <- rownames(fit$loadings)
rownames(fit$loadings) <- c(1:ncol(R))
Ident <- as.data.frame(cbind(c(1:ncol(R)), Names, Loading))
colnames(Ident) <- c("ID","HUGO","Loading on")

pdf("Identifiers.pdf", height = 23)
grid.table(Ident)
dev.off()

dandpal <- rev(rainbow(100, start = 0.4, end = 0.6))
dandelion(fit$loadings, bound = .3, mcex = c(1,1), palet = dandpal)
\end{lstlisting}

\bigskip
The code in Listing \ref{LIST:Bench} was used in the benchmark exercise contained in Section 3 of this supplement:

\begin{lstlisting}[language = R, caption = Code used in benchmarking, label = LIST:Bench]
####################################################
##--------------------------------------------------
## Benchmark Cn-plot and other methods
##--------------------------------------------------
####################################################

S <- c(125,250,500,1000)
p <- c(125,250,500,1000)

####################################################
## Alternative ridge estimator (equation 3)
####################################################

## Rotation equivariant setting
seed <- 1234
for (i in 1:length(S)){
  for (j in 1:length(p)){
    ## Generate data
    su = S[i]
    pu = p[j]
    nu = 200
    set.seed(seed)
    Y = matrix(rnorm(nu*pu), nrow = nu, ncol = pu)
    Target <- default.target(cor(Y), type = "DUPV")
    Sy <- cor(Y)

    ## Benchmark
    tm <- microbenchmark(CNplot(Sy, .00001, 20,
                                step = su, type = "Alt",
                                target = Target,
                                verbose = FALSE),
                         times = 50L)

    ## Save
    tm$expr <- mapvalues(tm$expr,
                         from = c(levels(tm$expr)[1]),
                         to = c("Condition number plot"))
    save(tm, file = paste("Alt.BM.RE.S",su,"p",pu,".Rdata", sep = ""))

    ## Plot
    boxplot(tm, log = FALSE)
  }
}

## Rotation non-equivariant setting
seed <- 5678
for (i in 1:length(S)){
  for (j in 1:length(p)){
    ## Generate data
    su = S[i]
    pu = p[j]
    nu = 200
    set.seed(seed)
    Y = matrix(rnorm(nu*pu), nrow = nu, ncol = pu)
    Target <- default.target(cor(Y), type = "DUPV")
    Target[1,1] <- 2
    Sy <- cor(Y)

    ## Benchmark
    tm <- microbenchmark(CNplot(Sy, .00001, 20,
                                step = su, type = "Alt",
                                target = Target,
                                verbose = FALSE),
                         times = 50L)

    ## Save
    tm$expr <- mapvalues(tm$expr,
                         from = c(levels(tm$expr)[1]),
                         to = c("Condition number plot"))
    save(tm, file = paste("Alt.BM.RNE.S",su,"p",pu,".Rdata", sep = ""))

    ## Plot
    boxplot(tm, log = FALSE)
  }
}


####################################################
## Archetypal Type I ridge estimator (equation 1)
####################################################

## Rotation equivariant setting
seed <- 9101112
for (i in 1:length(S)){
  for (j in 1:length(p)){
    ## Generate data
    su = S[i]
    pu = p[j]
    nu = 200
    set.seed(seed)
    Y = matrix(rnorm(nu*pu), nrow = nu, ncol = pu)
    Target <- default.target(cor(Y), type = "DUPV")
    Sy <- cor(Y)

    ## Benchmark
    tm <- microbenchmark(CNplot(Sy, .00001, 1,
                                step = su, type = "ArchI",
                                target = Target,
                                verbose = FALSE),
                         times = 50L)

    ## Save
    tm$expr <- mapvalues(tm$expr,
                         from = c(levels(tm$expr)[1]),
                         to = c("Condition number plot"))
    save(tm, file = paste("ArchI.BM.RE.S",su,"p",pu,".Rdata",
         sep = ""))

    ## Plot
    boxplot(tm, log = FALSE)
  }
}

## Rotation non-equivariant setting
seed <- 13141516
for (i in 1:length(S)){
  for (j in 1:length(p)){
    ## Generate data
    su = S[i]
    pu = p[j]
    nu = 200
    set.seed(seed)
    Y = matrix(rnorm(nu*pu), nrow = nu, ncol = pu)
    Target <- default.target(cor(Y), type = "DUPV")
    Target[1,1] <- 2
    Sy <- cor(Y)


    ## Benchmark
    tm <- microbenchmark(CNplot(Sy, .00001, 1,
                                step = su, type = "ArchI",
                                target = Target,
                                verbose = FALSE),
                         times = 50L)

    ## Save
    tm$expr <- mapvalues(tm$expr,
                         from = c(levels(tm$expr)[1]),
                         to = c("Condition number plot"))
    save(tm, file = paste("ArchI.BM.RNE.S",su,"p",pu,".Rdata",
         sep = ""))

    ## Plot
    boxplot(tm, log = FALSE)
  }
}


####################################################
## Archetypal Type II ridge estimator (equation 2)
####################################################

seed <- 17181920
for (i in 1:length(S)){
  for (j in 1:length(p)){
    ## Generate data
    su = S[i]
    pu = p[j]
    nu = 200
    set.seed(seed)
    Y = matrix(rnorm(nu*pu), nrow = nu, ncol = pu)
    Sy <- cor(Y)

    ## Benchmark
    tm <- microbenchmark(CNplot(Sy, .00001, 20,
                                step = su, type = "ArchII",
                                verbose = FALSE),
                         times = 50L)

    ## Save
    tm$expr <- mapvalues(tm$expr,
                         from = c(levels(tm$expr)[1]),
                         to = c("Condition number plot"))
    save(tm, file = paste("ArchII.BM.RE.S",su,"p",pu,".Rdata",
         sep = ""))

    ## Plot
    boxplot(tm, log = FALSE)
  }
}


####################################################
## Root-finding LOOCV and approximate LOOCV
####################################################

n <- c(100,200)
p <- c(125,250)
S <- c(125,250)

## Root-finding LOOCV, equivariant
seed <- 90210
for (i in 1:length(n)){
  for (j in 1:length(p)){
    ## Generate data
    pu = p[j]
    nu = n[i]
    set.seed(seed)
    Y = matrix(rnorm(nu*pu), nrow = nu, ncol = pu)
    Target <- default.target(cor(Y), type = "DUPV")

    ## Benchmark
    tm <- microbenchmark(optPenalty.LOOCVauto(Y, .00001, 20,
                                              type = "Alt",
                                              target = Target),
                         times = 50L)

    ## Save
    tm$expr <- mapvalues(tm$expr,
                         from = c(levels(tm$expr)[1]),
                         to = c("Root-finding LOOCV"))
    save(tm, file = paste("rfLOOCV.BM.RE.n",nu,"p",pu,".Rdata",
         sep = ""))

    ## Plot
    boxplot(tm, log = FALSE)
  }
}

## Root-finding LOOCV, non-equivariant
seed <- 902102
for (i in 1:length(n)){
  for (j in 1:length(p)){
    ## Generate data
    pu = p[j]
    nu = n[i]
    set.seed(seed)
    Y = matrix(rnorm(nu*pu), nrow = nu, ncol = pu)
    Target <- default.target(cor(Y), type = "DUPV")
    Target[1,1] <- 2

    ## Benchmark
    tm <- microbenchmark(optPenalty.LOOCVauto(Y, .00001, 20,
                                              type = "Alt",
                                              target = Target),
                         times = 50L)

    ## Save
    tm$expr <- mapvalues(tm$expr,
                         from = c(levels(tm$expr)[1]),
                         to = c("Root-finding LOOCV"))
    save(tm, file = paste("rfLOOCV.BM.RNE.n",nu,"p",pu,".Rdata",
         sep = ""))

    ## Plot
    boxplot(tm, log = FALSE)
  }
}

## Approximate LOOCV, equivariant
seed <- 902103
for (i in 1:length(n)){
  for (j in 1:length(p)){
    for(k in 1:length(S)){
      ## Generate data
      pu = p[j]
      nu = n[i]
      su = S[k]
      set.seed(seed)
      Y = matrix(rnorm(nu*pu), nrow = nu, ncol = pu)
      Target <- default.target(cor(Y), type = "DUPV")

      ## Benchmark
      tm <- microbenchmark(optPenalty.aLOOCV(Y, .00001, 20, step = su,
                                             type = "Alt",
                                             target = Target,
                                             verbose = FALSE),
                          times = 50L)

      ## Save
      tm$expr <- mapvalues(tm$expr,
                           from = c(levels(tm$expr)[1]),
                           to = c("Approximate LOOCV"))
      save(tm, file = paste("aLOOCV.BM.RE.n",nu,"p",pu,"S",su,
           ".Rdata", sep = ""))

      ## Plot
      boxplot(tm, log = FALSE)
    }
  }
}
\end{lstlisting}


\addresseshere

\end{document}